\pdfoutput=1

\documentclass[letterpaper,twocolumn,10pt]{article}
\usepackage{epsfig,endnotes,appendix}
\usepackage{color}
\usepackage[hyphens]{url}
\usepackage{hyperref}
\usepackage[numbers,sort]{natbib}

\def\showall{1} 

\def\reportversion{1.0}
\def\reportdate{May 8, 2020}

\begin{document}

\date{}

\title{\Large \bf COVID-19 Contact Tracing and Privacy: \\ Studying Opinion and Preferences} 

\author{
{\rm Lucy Simko$^{1,2,3}$,
Ryan Calo$^{2,4}$,
Franziska Roesner$^{1,2,3}$,
Tadayoshi Kohno$^{1,2,3}$}
\\
\\
    $^1$ Security and Privacy Research Lab, University of Washington
\\
    $^2$ Tech Policy Lab, University of Washington
\\
    $^3$ Paul G. Allen School of Computer Science \& Engineering, University of Washington
\\
    $^4$ School of Law, University of Washington
\\
Report Version \reportversion, \reportdate
\\
\url{https://seclab.cs.washington.edu/covid19}
} 

\maketitle

\thispagestyle{empty}

\textbf{This report (version 1.0) is deprecated: see arXiv:2012.01553 for our most recent report (version 2.0), published December 4, 2020, containing measurements through November 2020.}

\begin{abstract} There is growing interest in technology-enabled contact tracing, the process of identifying potentially infected COVID-19 patients by notifying all recent contacts of an infected person. Governments, technology companies, and research groups alike recognize the potential for smartphones, IoT devices, and wearable technology to automatically track ``close contacts'' and identify prior contacts in the event of an individual's positive test. However, there is currently significant public discussion about the tensions between effective technology-based contact tracing and the privacy of individuals. To inform this discussion, we present the results of a sequence of online surveys focused on contact tracing and privacy, each with 100 participants. Our first surveys were on April 1 and 3, and we report primarily on those first two surveys, though we present initial findings from later survey dates as well.  Our results present the diversity of public opinion and can inform the public discussion on whether and how to leverage technology to reduce the spread of COVID-19. We are continuing to conduct longitudinal measurements, and will update this report over time; citations to this version of the report should reference Report Version \reportversion, \reportdate. 

\end{abstract}

\section{Introduction}

Technology companies, university research groups, and governments are rapidly working to develop and deploy contact tracing apps to track and mitigate the spread of COVID-19. 
Prior work has determined that contact tracing apps will be most effective when used by the majority of a population~\cite{forbesContactTracing, ferretti2020quantifying}; however, some have raised security and privacy concerns (e.g.,~\cite{acluPrivacyConcerns, effProximityAppsChallenges}) as well as broader concerns about efficacy (e.g.,~\cite{efficacyCTapps}). 

Our research seeks to provide the scientific, technology, and policy communities with a better informed understanding of the public's privacy values, concerns, and opinions around the use of proposed automated contact tracing technologies. We argue neither for nor against automated contact tracing in this work, but instead offer a summary of public opinion on potential contact tracing scenarios, as we observe that many regions are moving towards automated contact tracing programs. We ask the following research questions:
\begin{itemize}
    \item \textbf{Data sources.} What data sources do people feel most and least comfortable with being used for contact tracing? We ask about multiple data sources, including: cell tower data, data from an existing app, data from a new app by a known entity or company, proximity data, credit card history, and surveillance camera footage.\footnote{We added in the questions about proximity data, credit card history, and surveillance camera footage questions after our initial survey. These questions do not appear in our first two weeks of survey data, which focus on location data.}
    \item \textbf{Data sharing and usage} What are people's opinions regarding contact tracing data being shared with or used by different entities for the purposes of contact tracing? We ask about data sharing with and usage by multiple entities, including: their government, cellular provider or cellphone manufacturer, and various well-known technology companies.
    \item \textbf{Other privacy concerns and mitigations.} What other privacy concerns do people have when deciding whether to download a contact tracing app? In what circumstances are people most likely to download and use a contact tracing app? 
\end{itemize}

We capture public opinion through online surveys using an international paid survey platform (Prolific). Our first two surveys were on April 1 and 3, 2020, and we are continuing to repeat variants of this online survey at regular intervals. Each survey is with 100 participants. Our April 1 and 3 surveys were from  before the virus had reached its peak infection rate, before contact tracing apps were ubiquitously available in many countries, and early in the public discourse about contact tracing. This version of our paper focuses on the April 1 and 3 survey results, though we do also discuss initial findings from later surveys as well.

We are releasing our results now, although preliminary, so that they can inform (1) ongoing technical efforts to design contact tracing apps in a privacy-preserving manner, (2) how the makers of such a contact tracing app or program communicate the privacy properties of their contact tracing program to their potential users, and (3) legal and policy discussions around the appropriate use of such technologies. 

To sample some preliminary findings:
\begin{enumerate}
    \item From April 1 and 3, 2020 (week 1): if a contact tracing app ``protected [their] data perfectly'', 72\% of participants said they would be at least somewhat likely to download it; as privacy risks increase, their self-reported download likelihood decreases (Section~\ref{section:likelihood}).
    \item From April 1 and 3, 2020 (week 1): participants preferred a company that they already trust with respect to security and privacy and that they perceive to have the resources and domain knowledge to add location tracking and analysis. To some, it is important that location tracking is already the primary purpose of the app, if added to an existing app. Many expressed a more positive view of specific companies (e.g., Google) or apps (e.g., Google Maps), even before Google and Apple announced their efforts (Sections~\ref{section:existing-app},~\ref{section:who_develops}).
    \item From April 1 and 3, 2020 (week 1): support for government use of location data (for the purposes of studying or mitigating COVID-19) varied by situation, but was never strongly supported. Participants were more comfortable than not with their government analyzing their cell tower location data if they tested positive for COVID-19, but also expressed a strong lack of confidence that their government would use the data only for COVID-19 mitigation; over half were concerned that data
    sharing with their government would bring harm to themselves or their community (Sections~\ref{section:cell-tower},~\ref{section:who_develops},~\ref{section:gov_use}). 
    \item From April 17, 23, and May 1 (weeks 3-5): participants may not view proximity tracking as more privacy-respecting than location tracking. Participants reported, through separate questions about location tracking apps and proximity tracking apps, they would be no more likely to download a proximity tracking app than they would a location tracking app in which the app makers knew their location. They indicated concerns about the security and privacy of proximity tracking that may not be consistent with the proximity tracking developers' threat models. We encourage contact tracing and proximity tracking developers to consider how to best communicate the threat model that their protocol defends against to non-technical users, and whether  that  threat  model  matches  up  with  what their  users  actually  want  and  need (Section~\ref{section:later-results}). 
\end{enumerate}

Stepping back, we stress that this report primarily documents findings from our April 1 and 3, 2020 surveys, though we do discuss some results from later survey dates. Public sentiment may change over time, as the number of infected people changes, as new technologies emerge, as new policies are proposed or enacted, and as new public discussions ensue. A future version of this paper will report on our longitudinal measurements. Additionally, we stress that our study focuses on the interplay between privacy and contact tracing, and does not consider the numerous other user-centered challenges around automated contact tracing, such as algorithmic accountability or differences in the availability and usability of smartphones or apps across populations. We surface these questions further in Section~\ref{section:broader} and urge other researchers, app developers, technology companies, and governments to consider these issues as they consider developing and deploying automated contact tracing programs.

\section{Limitations}

We are presenting these preliminary results without peer review because we believe our results have the potential to inform the current rapidly-evolving technical work on contact tracing apps.  However, because the paper has yet to undergo peer review, and despite our good faith attempts never to mislead or misrepresent our findings, we hope the reader will be especially cautious in interpreting our results.

We also stress that these results are \textit{preliminary}: we present the raw percentages of participants who answered each multiple choice question, along with \textit{N}, the number of participants who answered that question (sample size). We have not done statistical analysis of variance within questions or correlation across questions or participant groups. We will update this paper with additional analyses, including our analysis of longitudinal data and statistical analyses, over time.

Finally, online surveys such as ours have inherent limitations. Participants may experience survey fatigue and click through long matrix questions, giving inaccurate answers in order to finish the survey more quickly. From initial qualitative analysis, responses to free response questions seem to be on topic and high quality, indicating a low rate of survey fatigue. Survey fatigue, or any lack thereof, may also be affected by the fact that participants were payed, and therefore incentivized to finish.

Prior work on Mechanical Turk participants in the United States --- a different survey platform than the one we used --- has found, with varied results, that online survey participants may not be representative of the general population~\cite{ross2010crowdworkers}. Other studies have examined whether online survey participants' security and privacy knowledge and behavior accurately represent the general public, with varying results~\cite{kang2014privacy, redmiles2019well}.

\section{Location tracking during the COVID-19 pandemic}

\begin{table}
    \centering
    \begin{tabular}{|c|c|c|c|c|}
    \hline
    Country & Total pop & 4/1 cases & 5/1 cases \\  
    \hline 
    UK & 65.8m & 29.9k & 178.7k  \\
    \hline
    USA & 332.6m & 213.2k & 1.1m  \\
    \hline
    Europe & 741.m & 429k & 1.4m \\
    \hline
    \end{tabular}
    \caption{State of the world on April 1. European countries represented in our survey on April 1 + 3 are: Portugal ($\le$15\% of participants); Poland ($\le$10\% of participants); Greece, Italy, Mexico, Netherlands, Spain, Hungary, New Zealand, Germany, Finland, Israel, Estonia, Denmark, Slovenia, Turkey, Latvia, Belgium, and France ($\le$5\% of participants). Other countries represented in our survey on April 1 + 3 are: Canada, Mexico, Israel, Chile, Australia ($\le$5\% of participants)~\cite{jhu, europeApril1Covidnumbers, ciaWorldFactbook}.  
    }
    \label{fig:april1}
\end{table}

Because the pace of development on contact tracing apps is so rapid, and because the pandemic situation is rapidly evolving around the world, we expect this section to be quickly outdated. However, it is important to capture the state of the world on April 1, when we first deployed our survey (and the results of which we report on here), and how contact tracing efforts have grown since then, in order to contextualize our results. 

\subsection{State of the world as of April 1}

On April 1, the course of the COVID-19 pandemic had not yet peaked outside Asia. Table~\ref{fig:april1} shows the number of infections on April 1 and May 1 in the top three broad regions in our survey, the UK, US, and Europe (minus the UK), representing 88.5\% of survey participants. As of this writing, many European countries---e.g., Spain, Italy, and Germany---have seen slowing or slightly declining rates of infection~\cite{europeDelineRate}. 

On April 1, many European countries, including the UK, Germany, Italy, and Spain, were under varying forms of lock down, with some combination of schools, restaurants, bars, and non-essential shops closed, public gatherings banned, and citizens urged or mandated to stay inside except for essential outings~\cite{germanyContractTracingApril1, march31, march30Spain, ukLockdown}. Many in the US were under similar restrictions, though some states issued state-wide mandates after April 1, and other states did not issue a stay-at-home order (and as of this writing in early May, some states are lifting or planning to lift restrictions)~\cite{April1NYT, usStatesApril1, usStatesApril1_2, nytSeeWhichStates}.

To our knowledge --- and our participants' knowledge --- contact tracing apps were not available in most of the countries from which we drew participants as of April 3. Only 9 of 200 participants indicated that they had downloaded an app whose purpose was studying or mitigating COVID-19, and additionally 3 indicated that there was an app available that they had not downloaded. All others said that either an app was not available or they were not sure if an app was available.

\subsection{Contact tracing efforts}

Here we give a brief overview of existing contact tracing app efforts and the conversation around how to contact trace in a privacy-preserving way, captured at this point in time (early May). The purpose of this section is to contextualize our findings and recommendations, not give a comprehensive look at automated or technology-enabled contact tracing efforts.  

\paragraph{Why automated contact tracing.} Traditionally, contact tracing is done by a team of public health experts and focuses on tracking down those who might have been infected by someone who tested positive for a disease, in combination with widespread testing. There are multiple reasons that a state, region, or other entity might implement or endorse automated contact tracing (likely to augment or complement human-based efforts) though not all experts agree that automated contact tracing is needed or will be effective. For example, as areas move toward stopping ``stay at home'' orders, automated contact tracing might be used to keep the rate of infection low while allowing people to leave the home. Contact tracing technologies might also be used to enforce quarantine for people who have been identified as COVID-19-positive.

\paragraph{Existing automated contact tracing programs as of early May.}

Some governments have already deployed contact tracing apps or programs, using a variety of devices and data sources~\cite{wikipediaCovidApps}. For example, contact tracing apps exist in Bahrain, China, Colombia, the Czech Republic, Ghana, India, Israel, the Republic of North Macedonia, Norway, Singapore, and some US States~\cite{chinaContactTracing, traceTogether, bahrainApp, colombiaApp, czechApp, ghanaApp, indiaApp, israelContactTracing, israelApp, macedoniaApp, norwayApp, utahContactTracing, northDakotaContactTracing}. Some apps are mandatory (e.g., in China), and most are optional (e.g.,  Singapore) but may struggle with low adoption~\cite{singaporeDownloadRate}. 
Hong Kong deploys electronic wristbands to those infected with COVID-19 in order to ensure they were not leaving their homes~\cite{hongKongWristbands}. In South Korea, the government sends text messages to everyone in a region with the details of new COVID-19 cases, and there is a central database available with anonymized information; however, some entries in the database have been specific enough to be traced back to a single person, and have started damaging rumors~\cite{southKoreaContactTracing, southKoreaCoronaMap}. Taiwan and Israel both use cell tower data:  Taiwan uses it for for quarantine geofencing, while Israel uses it for contact tracing (Israel's data comes from a previously secret database used for counterterrorism)~\cite{israelContactTracing, taiwanGeofencing}. There may also be programs in other regions; this is not meant to be an exhaustive list. 

There is also significant interest in exploring automated contact tracing by other countries and regions, and growing efforts by technology companies, startups, governments, and universities to develop such programs~\cite{unifiedDoc}.

Additionally, several companies have leveraged geolocation data to create ``mobility reports'' that measure the extent of social distancing in specific areas, but do not identify or notify individual users~\cite{appleMobility, googleMobilityReport, cuebiq, unacast}. 

\paragraph{Design decisions affecting security and privacy.}

There are design properties at multiple levels that affect user security and privacy, some of which users may be inherently aware of (e.g., being potentially identified as infectious in some designs), and some of which are abstracted more from the user (e.g., broadcast vs narrowcast; centralization vs decentralization). For a more complete and in-depth discussion of these properties, see~\cite{redmilesUserConcerns, centralizedOrDecentralized}. 

Some groups explicitly focusing on privacy-respecting contact tracing --- each of which is making design decisions based on their threat models and on-the-ground situations --- include Apple and Google, the Massachusetts Institute of Technology (MIT), the University of Washington (UW), PEPP-PT, Inria, and DP3T~\cite{appleGoogleAnnouncement, mitPact, uwPact, ROBERT, dp3t}. One high-level distinction that has gained more traction since our initial surveys in early April is proximity tracking, in which a user's phone tracks other nearby phones, rather than a more traditional implementation of contact tracing, in which a user's location is tracked.

Additionally, there are other, non-smartphone methods that are being used and discussed for future use in automated contact tracing, such as credit card purchase history, and facial recognition on surveillance camera footage~\cite{russiaFacialRecognition, singerAsCoronavirusNYT}.

\subsection{Broader issues and considerations.} \label{section:broader}

Beyond the privacy and security concerns and opinions that our work surfaces, there are many other broader issues that must be addressed before the release of a contact tracing application. Our work does not touch on these issues directly, but we would be remiss not to mention some of the broader impacts in a report about security and privacy. We present example issues below, but refer readers to ~\cite{equityContactTracing, equityContactTracing2, efficacyCTapps, effCivilLivertiesPublicHealth, redmilesUserConcerns, raskarAppsGoneRogue} for broader discussion of equity and efficacy concerns.

\paragraph{Availability for and usability by all.} Governments and app makers must consider the implications of an app being available only in certain languages, or without accessibility features. If a certain demographic group is left without access, or without usable access, they may experience different rates of infection. This lack of access may also lower the effectiveness of contact tracing. 

\paragraph{Users without smartphones.} Not all people have smartphones, and some high risk groups, such as seniors, may be less likely to regularly use a smartphone. A smartphone app disenfranchises those sets of users.

\paragraph{Mandatory vs optional use.} Some have raised concerns about the efficacy of optional contact tracing apps~\cite{efficacyCTapps}; however, mandatory apps raise different concerns about state surveillance, particularly if mandatory for only a certain demographic. Opt-in automated contact tracing efforts require the public to reason about the privacy trade-offs. Mandatory automated contact tracing efforts, meanwhile, could force some to use technology that runs counter to their privacy preferences (and potentially take purposefully evasive actions to avoid tracking). 

\paragraph{Explainability of contact tracing algorithm and psychological effects.} How should an app notify a user that they may be infected, and how much information should that app give about who infected them or when / where they were infected? How should an app communicate the expected false positive and false negative rate? If either false positives or false negatives are too high, users' mental health may suffer, or users may become overly confident that they are not infected.

\paragraph{Malicious actors.} One or more intentionally false reports of positive tests could have financial and social consequences for businesses or people who either believe they are infected, or whom other people believe are infectious.

\section{Related Work}

Other groups have also investigated public opinion on location tracking during COVID-19. Below we briefly describe three efforts, and note that although not all results match up exactly with respect to preferences from different demographic groups, each survey asked different questions, and on different dates.

Our results are also not exactly the same, but all seem to indicate (a) reservations about government location tracking from a significant portion of the population and (b) over half of the population is willing to download a contact tracing app under the right circumstances. (However, we note that ``the right circumstances'' may be different for different user groups, and that there is more work needed to determine how automated contact tracing might affect marginalized or under-resourced groups, and what needs those groups have with respect to contact tracing, if any).

The Pew Research Center, in a survey conducted April 7-12, reported that 60\% of Americans do not believe that location tracking by the government would mitigate the pandemic, and many have concerns about privacy or appropriateness of such tracking~\cite{pewMostAmericans}. They found that there are statistically significant differences between demographic groups delineated by political party, race/ethnicity, and age, with Republicans, Hispanic Americans, and those who are 30 or older being more accepting of the government tracking location for the purposes of mitigation COVID-19.

Hargittai and Redmiles surveyed around a thousand adults in the each of
Italy, Switzerland, and the US. They found that two thirds of the US participants would be willing to install a contact tracing app, ``that would help slow the spread of the virus and reduce the lockdown period, even if that app would collect information about their location data and health status''~\cite{hargittaiWillAmericans, hargittaiStudy}. They found that those in the US would most trust an app from a health agency like the CDC or their insurance provider over the other options they asked about, but that there is no single trusted source. Therefore, in a Scientific American blog, they recommend multiple applications that can interoperate, with a shared base technology, in order to allow users to pick the app that best suits their threat and usage model~\cite{hargittaiWillAmericans}.

In a survey conducted April 21-26, a group from The Washington Post and the University of Maryland found that 50\% of American smartphone users would not use an app using the contact tracing technology proposed by Apple and Google, and that a further 18\% of their respondents did not use a smartphone~\cite{wapoMostAmericans, wapoUmdSurveyProtocol}. The accompanying Washington Post article notes that smartphone usage rate is lower amongst seniors, who are more vulnerable to the disease. This survey found that Democrats would be more willing to download the app, and that many (57\%) Americans would trust the CDC to deploy such an app (consistent with Hargittai and Redmiles), but that fewer (47\%) would trust their health insurance company.

Redmiles, in a separate survey conducted May 1-5, found that around 80\% of Americans would be willing to download a contact tracing apps, depending on its efficacy, accuracy, and privacy~\cite{redmilesHowGood}.

Our work adds to these results: though we have fewer participants (and participants primarily from Europe), we are able to answer more nuanced questions about data sources, privacy preferences, and we also are able to distill some higher level user values and concerns from qualitative data. Our study is also longitudinal and positioned to observe changes in opinions over time.

\section{Methodology}

In order to collect rich data and measure public opinion, we designed an approximately 15-minute online survey with both multiple choice and free response questions. Our survey was implemented in Qualtrics. We deployed the survey through Prolific, an online survey platform based in the United Kingdom. 

Our institution's IRB determined our study was exempt from further human subjects review, and we followed best practices for ethical human subjects survey research, to the best of our knowledge, e.g., we paid at or slightly above minimum wage, all questions were optional except the initial questions about age and smartphone usage, and we did not collect unnecessary personal information. We did give participants the option to submit their email address in a separate Google form (not connected to their survey results) in case they wanted to be contacted about follow-up surveys.

\subsection{Survey Protocol}

Because we expected most participants to be from countries where an app to track and mitigate COVID-19 was not in ubiquitous use, at least at the time of our initial survey, most of the survey was designed to elicit user attitudes about contact tracing in specific \textit{hypothetical} situations, some more plausible than others (and some actually happening around the world at the time of the survey). The survey did include branches for those who had downloaded an app for tracking or mitigating COVID-19, or who had the opportunity to but chose not to; only a small minority of the participants followed these branches. In order to avoid biasing participants towards presenting themselves as more privacy-conscious than they are, the survey did not mention ``privacy'' until the final two questions (demographics), and asked instead about participants' ``comfort'' with various situations, or their ``likelihood'' of downloading an app in a certain situation.  Each section (except for demographics) concluded with one or more optional free-response questions, inviting participants to explain their answers.

The survey had the following main sections (excluding questions for participants who were already using a contact tracing app). 
\paragraph{Demographics.} We asked participants three types of demographic questions, centered around questions about COVID-19, or variables we hypothesized might correlate with their attitude towards COVID-19 and contact tracing programs: (1) standard demographic questions, like age, gender, geographic location; (2) general political views, news sources, and privacy and technology interest and knowledge (self-described); (3) COVID-19-specific questions, like their general level of concern about the pandemic, whether they live with someone who is in a high-risk group, and their beliefs and implementation about social distancing. We asked many of the demographic questions at the end of the survey in order to help mitigate stereotype threat.

\paragraph{Cell tower location data.} In this section, we asked participants how comfortable they were with their cell phone manufacturer or cellular carrier using their location data for the purposes of studying or mitigating the spread of COVID-19. We presented participants with three variants on the situation: their location data being shared with their government, their location data being shared with their government if they tested positive, and their location data being shared publicly if they tested positive.

\paragraph{Existing apps using GPS location data.} Next, we asked participants to imagine that ``the makers of an existing app on your phone started using your GPS location data to study or mitigate the spread of COVID-19.'' We chose 3 popular apps from each of 5 categories that we expected would use location data (navigation, social media, messaging, transportation, fitness) for a total of 15 apps. Participants rated their comfort with each of the 15 apps using their location data for mitigating the spread of COVID-19 on a 5-point Likert scale, with an additional option for ``I don't use this app.''

We then asked participants about their comfort level if one of the apps they regularly use, which might not be one of the ones in the list above, started using their location data to mitigate the spread of COVID-19. Next, we asked two free-response questions about the app that they regularly use that they would \textit{most} trust and the app that they would \textit{least} trust to study or mitigate COVID-19.

\paragraph{New app: perfect privacy.} We then asked participants to imagine a new app that would track their location at all times for the purposes of mitigating the spread of COVID-19, but that it would protect their data perfectly. We asked, on 5-point Likert scales, how likely they would be to install the app, and how it would change their current behavior.

\paragraph{New app: app makers know location at all times but don't share it.} Changing the previous scenario slightly, we asked participants to imagine a new app that would know their location at all times for the purposes of mitigating the spread of COVID-19, and this time the app makers would know their location at all time but would not share it. We again asked participants how likely they would be to download and use such an app. This time, we asked participants to rate their comfort with each of companies that made the same popular 15 apps we had shown them previously making this app. We expanded this list to include other companies in week 3 of the survey. We also asked about their comfort with five generic entities making such an app: a university research group, an activist group, an industry startup, your government, and the United Nations.

\paragraph{New app: app makers know location at all times and share with your government if you are diagnosed with COVID-19.} Changing the previous scenarios again, we asked participants this time about a situation in which the new app's makers share their location history with the government if they test positive for COVID-19. We asked, again, how likely they would be to download such an app, as well as their download likelihood in two variant situations: if the data were shared regardless of whether they tested positive, and if the government's use of the data were supervised by judge.

\paragraph{Other location data sources: Surveillance camera footage and credit card history.} Turning away from location data collected by an app, and in response to an evolving conversation about alternate data sources, starting in week 3 we asked participants next about their comfort with location history derived from surveillance camera footage and credit card purchase history. 

\paragraph{New app: Proximity tracking.}
Due to the growing discussions about and technical work on proximity tracking protocols and apps after April 1, in week 3 we added a group of questions about proximity tracking. We asked about proximity tracking by phone manufacturers, phone operating systems, a new app, and several well known companies or generic entities. 

\paragraph{Government use of location data.} In this section, we stepped back from scenarios about specific data sources and asked participants questions about a scenario in which their government acquires their location data for studying and mitigating COVID-19. We asked about their confidence in their government's deletion of the data after the pandemic, use of data for only COVID-19 tracking, and their general level of concern for their ``personal safety or the safety of those in their community.''

\paragraph{Desired features in a new COVID-19 mitigation app.} We then asked participants about a wide variety of roles that a potential new COVID-19 mitigation app might have in enforcing isolation or notifying people of potential infections, drawn from existing contract tracing apps or programs. For example, one feature we asked about was ``notify you if you came close to someone who later tested positive for COVID-19'' while another was ``automatically notify the authorities if people were not isolating as mandated.''

\paragraph{Location sharing with their government pre-pandemic.} Finally, we asked participants to rate their level of comfort with their location data being shared with their government in October 2019, i.e., before COVID-19. 
Since participants may not accurately recall their own previous beliefs, and may have been primed towards privacy-sensitivity by the rest of the survey, any results from this data must be treated with caution.

\subsection{Recruitment}

We recruited participants through Prolific, an online survey platform, with no demographic restrictions, since Prolific already requires that all participants be 18 or older. The first two questions of our survey screened participants as required by our IRB. We asked: (1) are you at least 18 years old? and (2) do you use a smartphone regularly?. If participants answered `Yes' to both, they proceeded to the rest of the survey.

We ran the survey on Prolific on April 1, 3, 8, 10, and every Friday thereafter, with 100 people each time, around the same time (3pm Pacific). We exclude participants who have taken any previous version of the survey. On the initial release of the survey, in early April, we also recruited participants using the snowballing technique through our own contacts on social media and over email. In this paper, we report primarily on the Prolific participants recruited on April 1 and 3, as well as cursory results of data collected on May 1; in future papers, we plan to include the participants recruited more organically, as well as deeper studies of the longitudinal data from the weekly deployments on Prolific.

\subsection{Analysis}

In this initial report, we present raw percentages, no statistical analysis. When presenting percentages, we give the number of participants who answered that question; although we recruited 100 participants for each survey run,\footnote{In week 3, although we recruited 100 participants, we actually ended up with 102, due to two participants finishing after the time allotted by Prolific, the survey platform through which we recruited} all questions were optional, and a small minority already had an app that tracked COVID-19 and thus did not take the main branch of the survey (we do not report on this small minority here).

\section{Results}

Here we report primarily on initial analysis of quantitative results from week 1 of our survey, with 100 Prolific participants on April 1 and April 3 each (200 total). For this initial report, our current analysis do not include rigorous statistical analysis, but instead report only on the percentages of users that give each response to a question. We present these initial results --- with caveats --- because the issue is timely.

9 of the 200 participants from April 1 and 3 followed a survey branch indicating they already had a COVID-19 tracking app; we do not report on these 9 participants here.

Two notes on terminology: we use the term `contact tracing' as a superset of `location tracking', and `proximity tracking'. When reporting qualitative results, we use the format W1P100 to mean participant 100 from week 1.

In the text below, our notation X[$N$], where $N$ is a number, refers to unique identifiers in the Qualtrics survey platform that we used. We use this notation for internal bookkeeping and question disambiguation as we continue to collect more data and revise this report over time.

\subsection{Participant Demographics}

Participants resided primarily in the UK (26\%), the US (16\%), 15 European countries (46.5\%), and 7 other countries (11.5\%). Our participants were mostly young, with nearly half being less than 25 years old, and 56\% male. Four participants identified as agender or genderfluid, and one participant identified openly as transgender and male. We manually bucketed gender identities as reported by participants in a free response text field; we believe we have stayed true to participants' gender identities when bucketing (e.g., by bucketing a `man' response as `male'), though we note that participants gender identities may change over time, and that our participants may have included more trans or gender non-conforming participants than disclosed as so. 
To avoid the effect of stereotype threat, we asked most demographic questions at the end of the survey.


\subsubsection{Participants concerned about COVID-19 and believe in social distancing}

\begin{figure}
    \includegraphics[scale=.55]{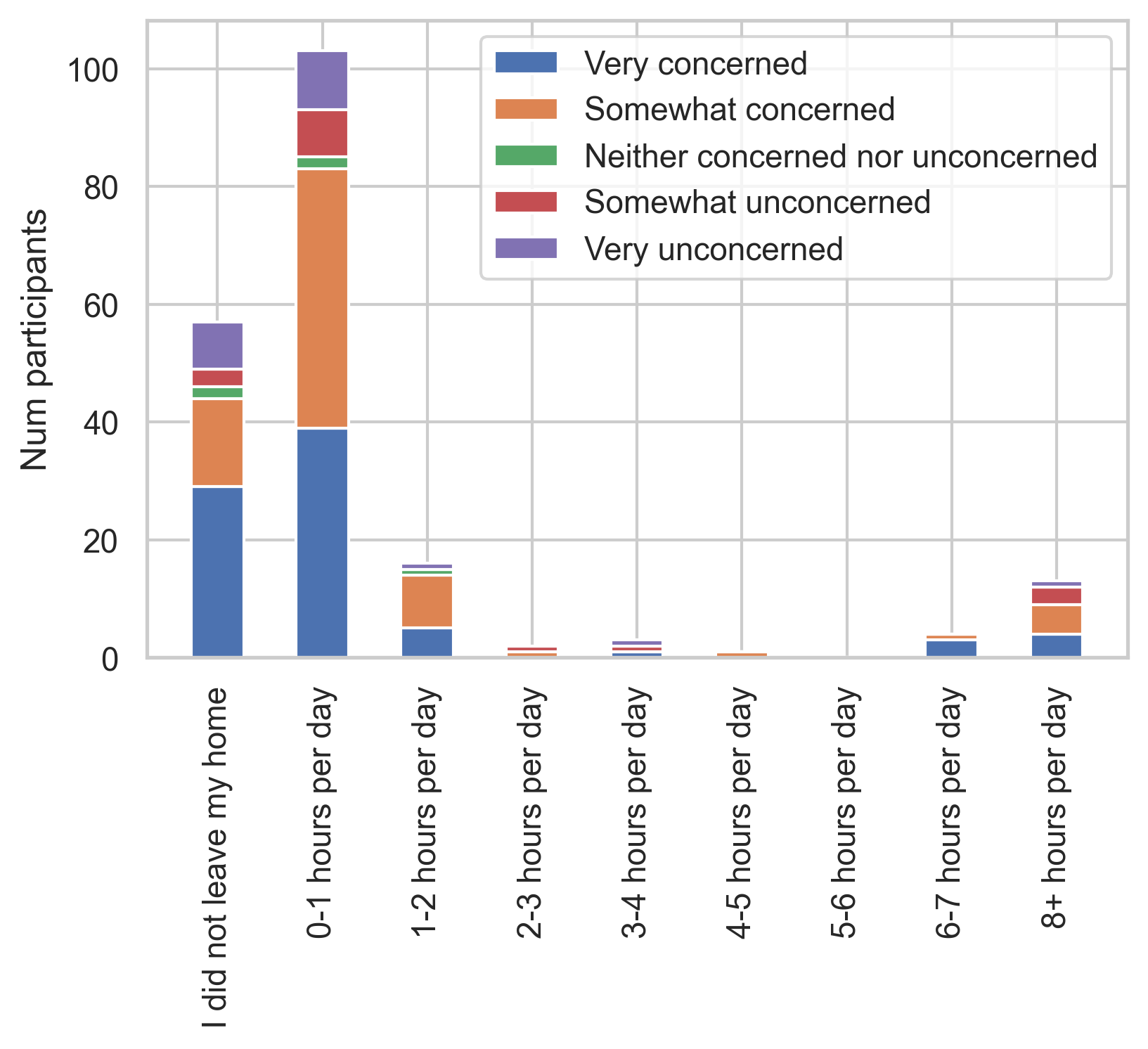}
    \caption{Participants' general concern about COVID-19 and the average time they spent away from home per day in the last week.}
    \label{fig:concern-hours}
\end{figure}

The majority of participants were concerned about COVID-19 (79\%, X6) and believed that social distancing is an important tool for slowing the spread of COVID-19 (96.5\%, X7). 80\% reported leaving their home less than one hour per day in the prior week, with 28.5\% not leaving their home at all (X8). However, a distinct minority (19\%) were unconcerned (X6) and a quarter spent 6 or more hours per day out of their home (though that is not necessarily at odds with social distancing) (X8), as shown in Figure~\ref{fig:concern-hours}

37\% of participants said they were either in a high risk group or living with someone considered high risk (X43). 

\subsection{Mostly positive attitude toward use of cell tower data for COVID-19 mitigation so long as it is not released publicly}
\label{section:cell-tower}

\begin{figure}
   \includegraphics[scale=.9]{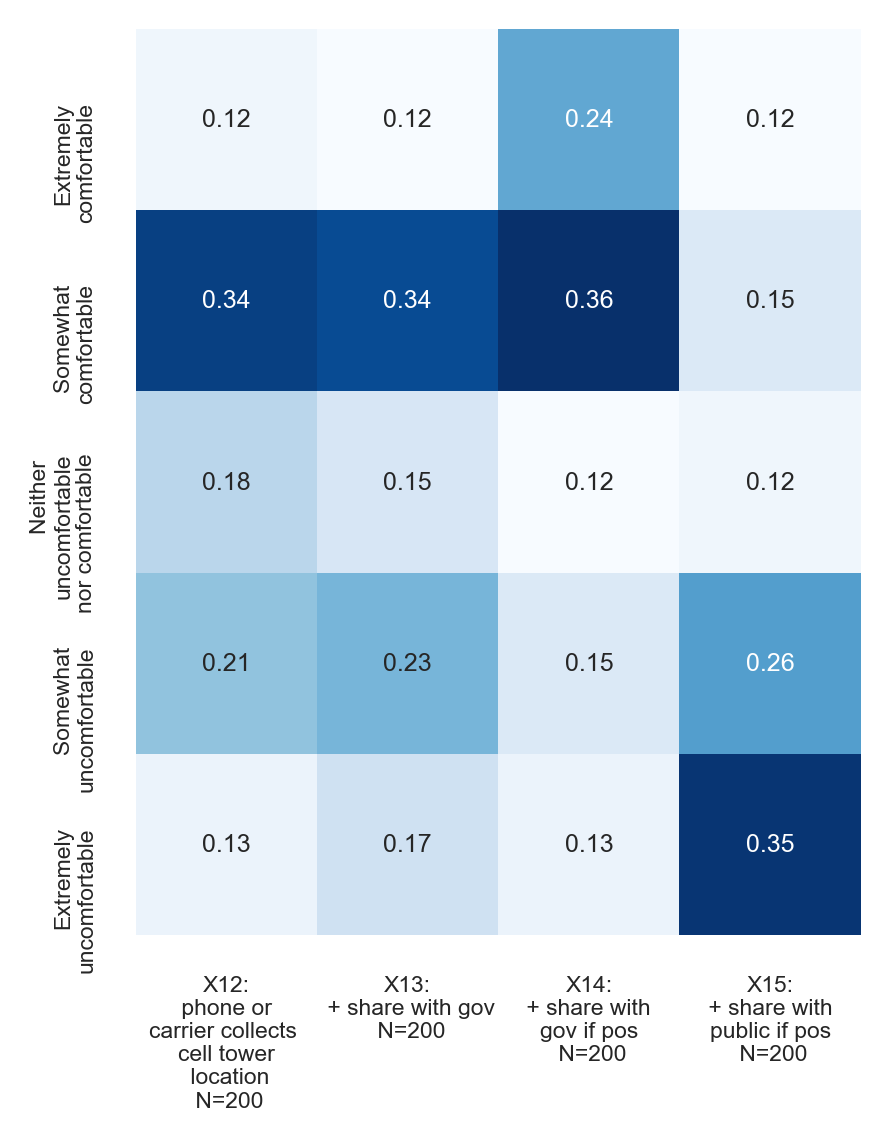}
    \caption{Attitudes towards cell tower location data being used for COVID-19 tracking. Columns sum to 1 and represent only the participants who answered}
    \label{fig:section2_0}
\end{figure}

Many participant were at least somewhat comfortable with their cell tower data being used for contact tracing; from the situations we presented, participants were most comfortable with their government being given the data only if they tested positive, and much less comfortable with the data being released publicly, as shown in Figure~\ref{fig:section2_0}. However, even amongst participants who were comfortable with their data being used (generically, as in X12, or shared with the government, as in X13 and X14), more were \textit{somewhat} comfortable rather than extremely comfortable, signifying their reservations. Even in the situation that participants were most comfortable --- the data being shared with the government if they tested positive ---  13\% were extremely uncomfortable, and a further 15\% were somewhat uncomfortable.

Of the situations presented, participants were most uncomfortable with the idea of their cell tower location history being shared publicly if they tested positive (61\% uncomfortable) (X15). This situation is reminiscent of South Korea's initial handling of location tracking: location data and biographical details were posted publicly and were not sufficiently anonymized, as groups discovered the identities of those who had tested positive and rumors started about extramarital affairs and plastic surgery trips~\cite{southKoreaRumors}. South Korea has since started anonymizing the publicly released data more thoroughly~\cite{singerAsCoronavirusNYT}.

This sentiment of being uncomfortable with public disclosure more generally underscores the importance of location data being properly protected when and if it is collected, to avoid data being made public or being exposed to other parties through a data breach.

\subsection{Amongst existing apps, participants are most comfortable with an already-trusted mapping app adding COVID-19 tracking}

\label{section:existing-app}

\begin{figure*}[t]
    \includegraphics{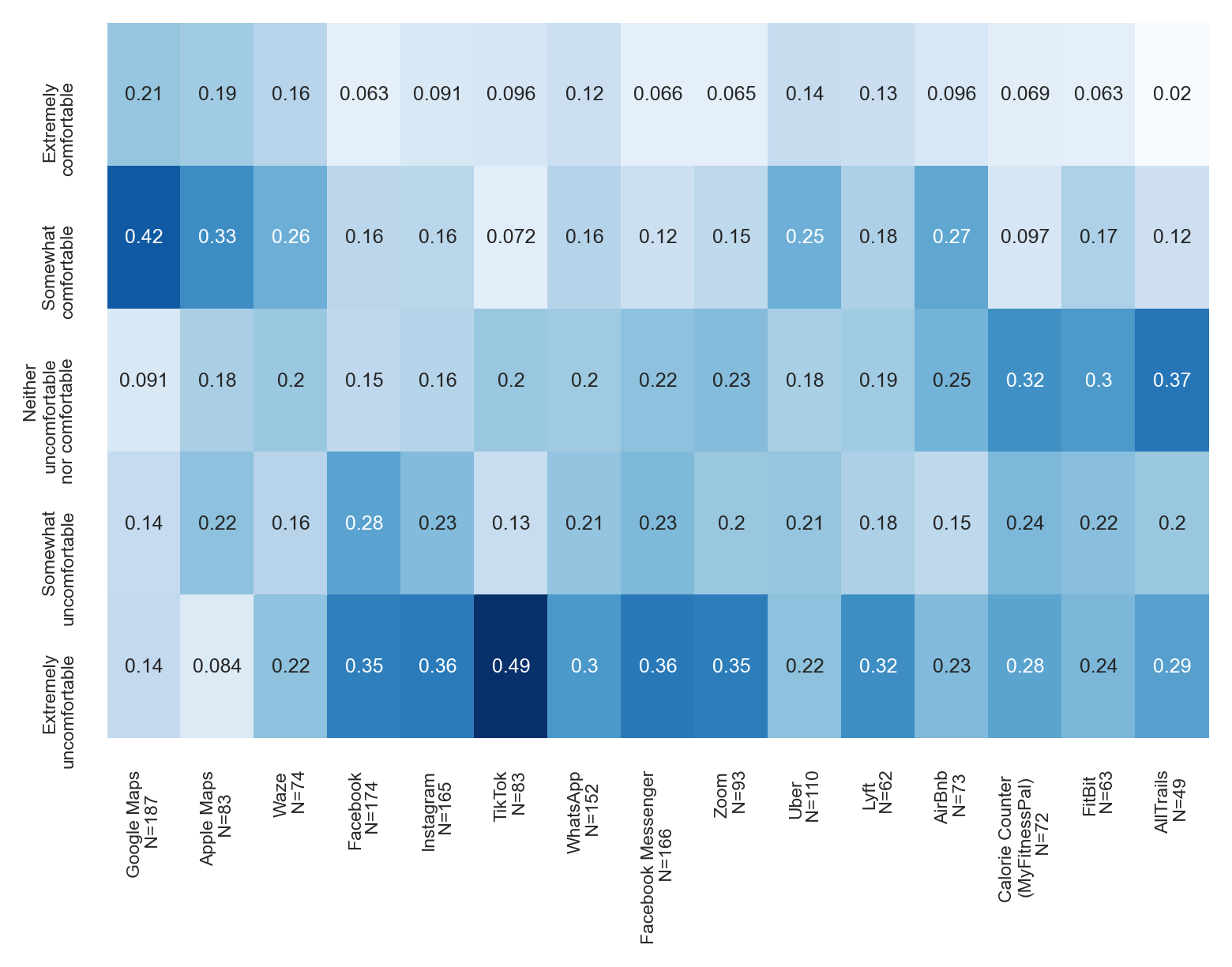}
    \caption{Participant comfort with an \textit{existing} app using their location to study or mitigate COVID-19 (numbers in columns add to 1 and are percentages; N = number of participants who use that app) (X18)}
    \label{fig:section2_05a}
\end{figure*}

Figure \ref{fig:section2_05a} shows a heatmap of user comfort with each of 15 apps hypothetically using their location data to study or mitigate COVID-19, showing only responses for participants who reported actually using the app. Participants were generally uncomfortable with the idea of most apps using their location data for COVID-19 tracking: 49\% would be extremely uncomfortable with TikTok adding this features; 30-36\% of participants would be extremely uncomfortable with any of the apps owned by Facebook that we asked about (Facebook, Instagram, WhatsApp and Messenger), and 21-28\% somewhat uncomfortable for the same apps.

Participants indicated the most comfort with Google Maps or Apple Maps using their location data to study or mitigate COVID-19 (among the 15 app options given): 63\% were at least somewhat comfortable with Google Maps, and 52\% were at least somewhat comfortable with Apple maps. This result was collected \textit{before} Apple and Google announced their collaboration on a contact tracing app on April 10, but user comfort reported on weeks 2-5 remained approximately the same (63-65\% at least somewhat comfortable with Google Maps; 49-54\% at least somewhat comfortable with Apple Maps)~\cite{appleGoogleAnnouncement}.

We note that some of these apps have a smaller sample size (e.g., TikTok, AllTrails), so that data should be interpreted with more caution. Additionally, there may be self-selection bias present in these results: if an app has a more privacy-conscious user base, its users may appear to be less comfortable with it using location data for COVID-19 tracking, and it may appear that that app is less trusted by all users. 

Preliminary analysis of qualitative results from week 1 in this section reveals common themes around user values and concerns regarding what apps they would trust most (X23) and least (X24) to use their location data for the purposes of COVID-19 tracking. In line with the quantitative results reported above, most users picked Google Maps as their most trusted app, and Facebook as their least trusted app, at least in the context of using location data for COVID-19 tracking. 

Participants' reasons for picking their most trusted app, or for not picking their least trusted app, include (a) the general reputation or public image, (b) a good history with privacy and security, (c) confidence in the company having the resources to add COVID-19 tracking to its services, and (d) location tracking already being central to the app's purpose.

Participants also mentioned several concerns, which extend to any contact tracing program: (a) stalking or personal harm due to poorly anonymized data; (b) data leakage or privacy breaches; (c) data being sold by the company; and (d) a ``slippery slope'' in which this sort of tracking becomes the norm.

\subsection{Privacy concerns reduce likelihood to install a new COVID-19 tracking app, from 72\% with perfect privacy}

\label{section:likelihood}
\begin{figure}
    \centering
    \includegraphics[scale=1]{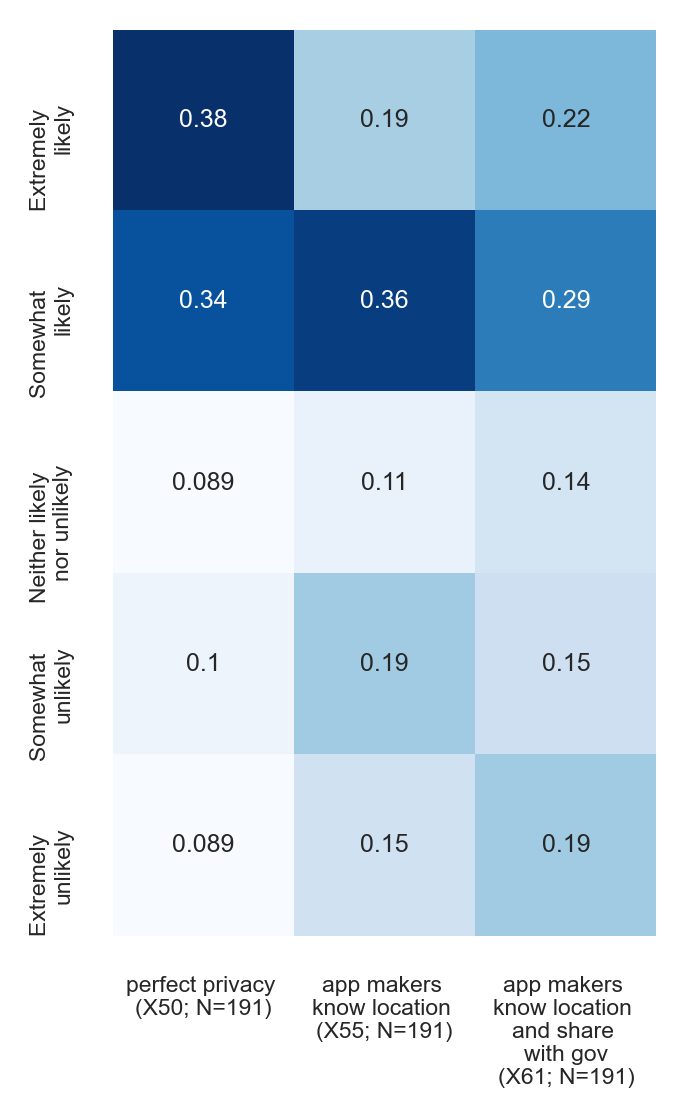}
    \caption{Self-reported likelihood of downloading a \textit{new} app to study or mitigate COVID-19, given three privacy situations: (X50) perfect privacy, (X55) app makers know location, (X61) app makers know location and share with the government if positive.}
    \label{fig:section3_0-how-likely}
\end{figure}

In response to three general data-sharing situations that might be present with a \textit{new} COVID-tracking app, participants indicated that in the ``best possible'' privacy situation --- ``an app that protects your data perfectly'' --- 72\% would be at least somewhat likely to download the app, as shown in Figure~\ref{fig:section3_0-how-likely}. Estimates vary on how much of a population needs to participate in contact tracing in order for it to be effective and depend other interventions, but recent work suggests a rate of around 60\%~\cite{ferretti2020quantifying, forbesContactTracing} to 70\%~\cite{hellewell2020feasibility}. 

If the app makers knew their location but did not share with anyone, participants were less comfortable: only 19\% would be extremely likely to download this app, and 36\% would be somewhat likely. Further, 35\% would be extremely or somewhat unlikely to download the app.

Finally, participants were least comfortable with app makers knowing their location and sharing it with their government, though more were still likely to download the app than unlikely: 49\% were at least somewhat likely, while 34\% were somewhat or extremely unlikely.

These situations show that even in the best possible situation, with ``an app that protects your data perfectly,''  most participants have reservations about using an app to study or mitigate COVID-19. It is worth noting that only about 20\% of Singaporeans had downloaded Singapore's contact tracing app, TraceTogether, as of April 1~\cite{nytDownloadRate, forbesContactTracing, singaporeDownloadRate}. 

Further, 74.8\% said that using such an app would cause no change in their behavior, 20.9\% said they would be less social, and 4.2\% reported that they would be more social (X51). Being less social while being tracked by an app could indicate that those 20.9\% of participants would feel uncomfortably surveilled when using the app  or that they think the app's notifications might alter their behavior, while the 4.2\% who would be more social might feel protected by such an app. 

Results from later surveys by others, and from our later surveys, show similar willingness to download. The Washington Post and University of Maryland effort found that 50\% of American smartphone owners would definitely or probably use a contact tracing app, while Hargittai and Redmiles found that about two-thirds of Americans would download such an app~\cite{wapoMostAmericans, hargittaiWillAmericans, hargittaiStudy}. Separately, Redmiles found that 82\% of Americans would be willing to download a contact tracing app in perfect privacy and accuracy conditions, but that their willingness to download would drop dramatically if data could leak~\cite{redmilesHowGood}.

\subsection{Participants care that app developers don't share data; are most comfortable with Google and the UN developing a new app}
\label{section:who_develops}

\begin{figure*}[t]
    \includegraphics[scale=1]{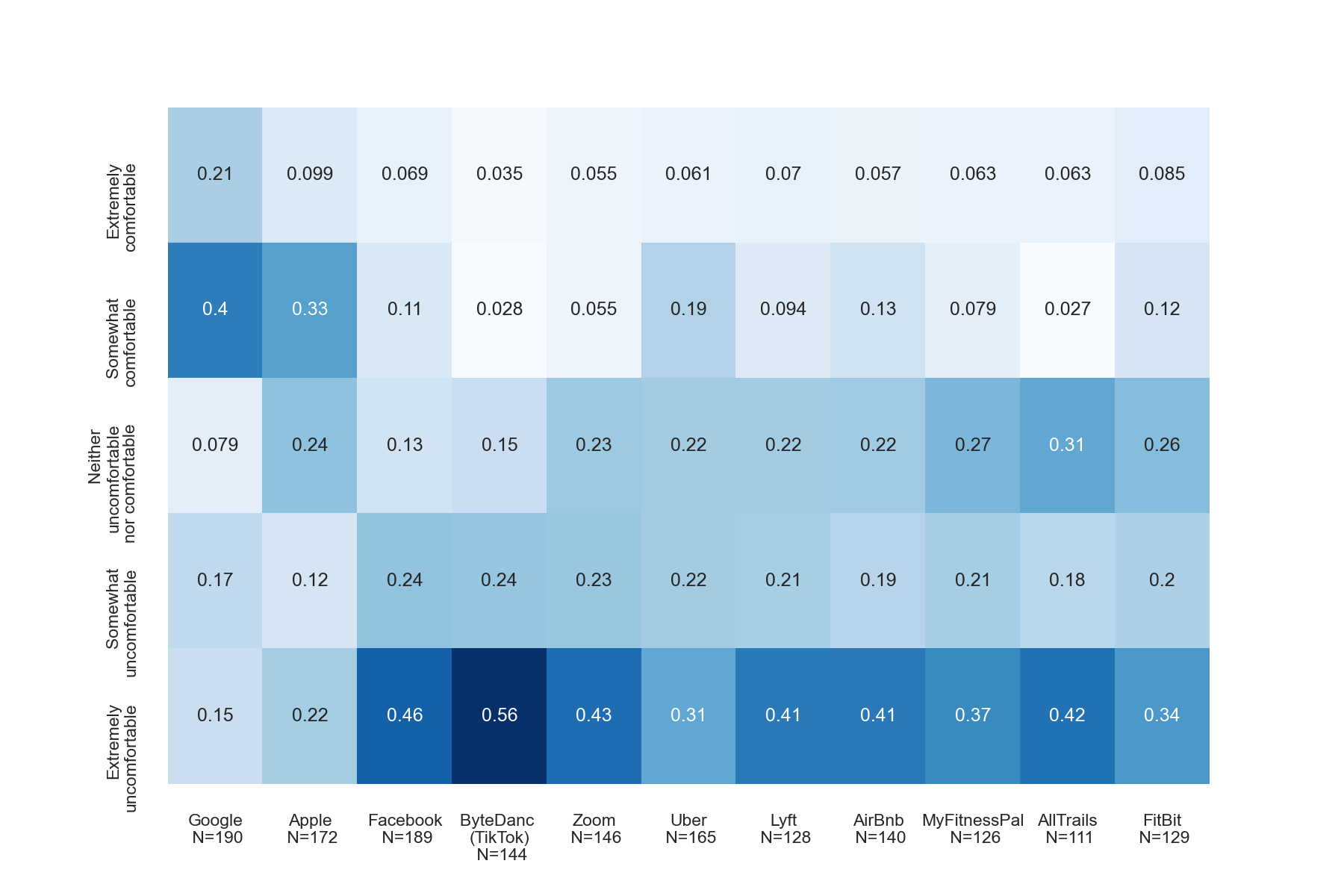}
    \caption{Participant comfort level with companies hypothetically creating a \textit{new} app to mitigate COVID-19, assuming that app would know their location at all times, but would not share the data (X56).}
    \label{fig:section3_1-companies}
\end{figure*}

Participant opinions about existing companies making a \textit{new} app followed similar trends as in Section~\ref{section:existing-app}, shown in Figure \ref{fig:section3_1-companies}: most participants were at least somewhat comfortable with Google (61\%), and many were somewhat comfortable with Apple (33\%; 10\% extremely comfortable), though just as many were uncomfortable with Apple (34\% somewhat or extremely uncomfortable). Participants were overwhelmingly uncomfortable with the other companies creating such an app. 

\begin{figure}
    \includegraphics[scale=.9]{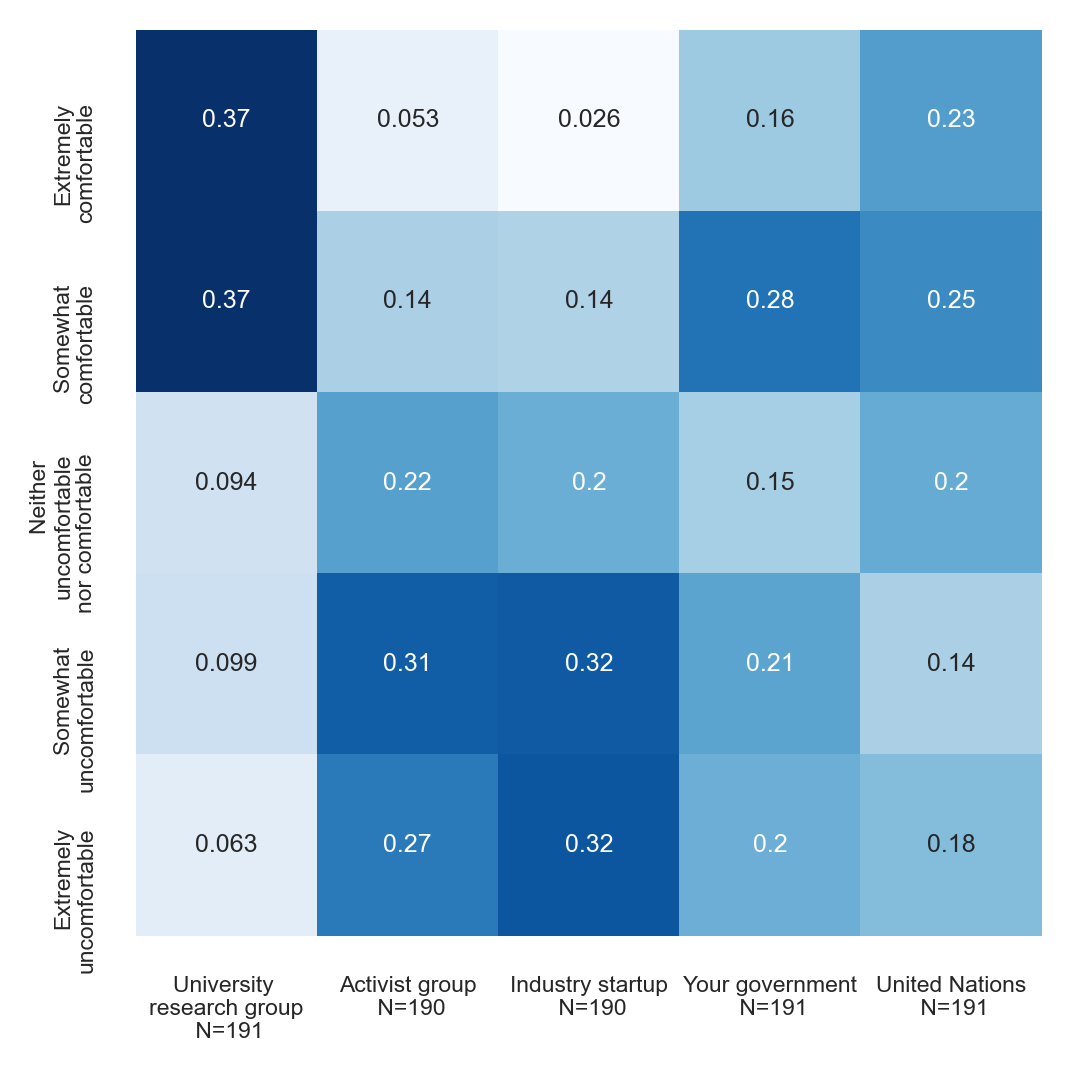}
    \caption{Participant comfort level with various generic entities creating a \textit{new} app to mitigate COVID-19, assuming that app would know their location at all times, but would not share the data (X57).}
    \label{fig:section3_1-entities}
\end{figure}

Participants also indicated general mistrust for a potential new COVID-19-tracking app created by an industry startup (64\% somewhat or extremely uncomfortable) or an activist group (58\% somewhat or extremely uncomfortable), as show in Figure~\ref{fig:section3_1-entities}. Responses indicate the most trust in a university-developed app (74\% somewhat or extremely comfortable), but we note that at the beginning of the survey, participants were shown our university's logo and told that this survey was an academic endeavour, which may have caused responses bias, as found by~\cite{dell2012yours}. Participants were largely split on generic trust for a government- or United Nations-developed app.

Initial analysis of qualitative responses to X58 echo the values and concerns reported in Section~\ref{section:existing-app}: participants were concerned about sharing and selling of their data, and lack of trust in multiple companies and activists and startups, with some perceived to have ulterior motives. W1P198 explained: ``\textit{if it saves lives then i would use the app if it was made by any company, i would be much happier if a university did it as they have little to gain by having my data, other than to use it for its intended reason.}'' Similar, W1P186 wrote: ``\textit{I[t] leaves me uncomfortable as I currently don't trust many companies for data protection, I could however be convinced if it were made by a trusted university or worldwide organization.}''

In contrast, Hargattai and Redmiles found that universities would be one of the least trusted entities, at less than 10\%, while the survey by the Washington Post and the University of Maryland found they were relatively trusted, at 57\%. This discrepancy highlights the need for multiple surveys and qualitative data to better understand the nuances of public opinion. 

\subsection{Unease over government use of data}
\label{section:gov_use}

\begin{figure}
    \includegraphics[scale=.6]{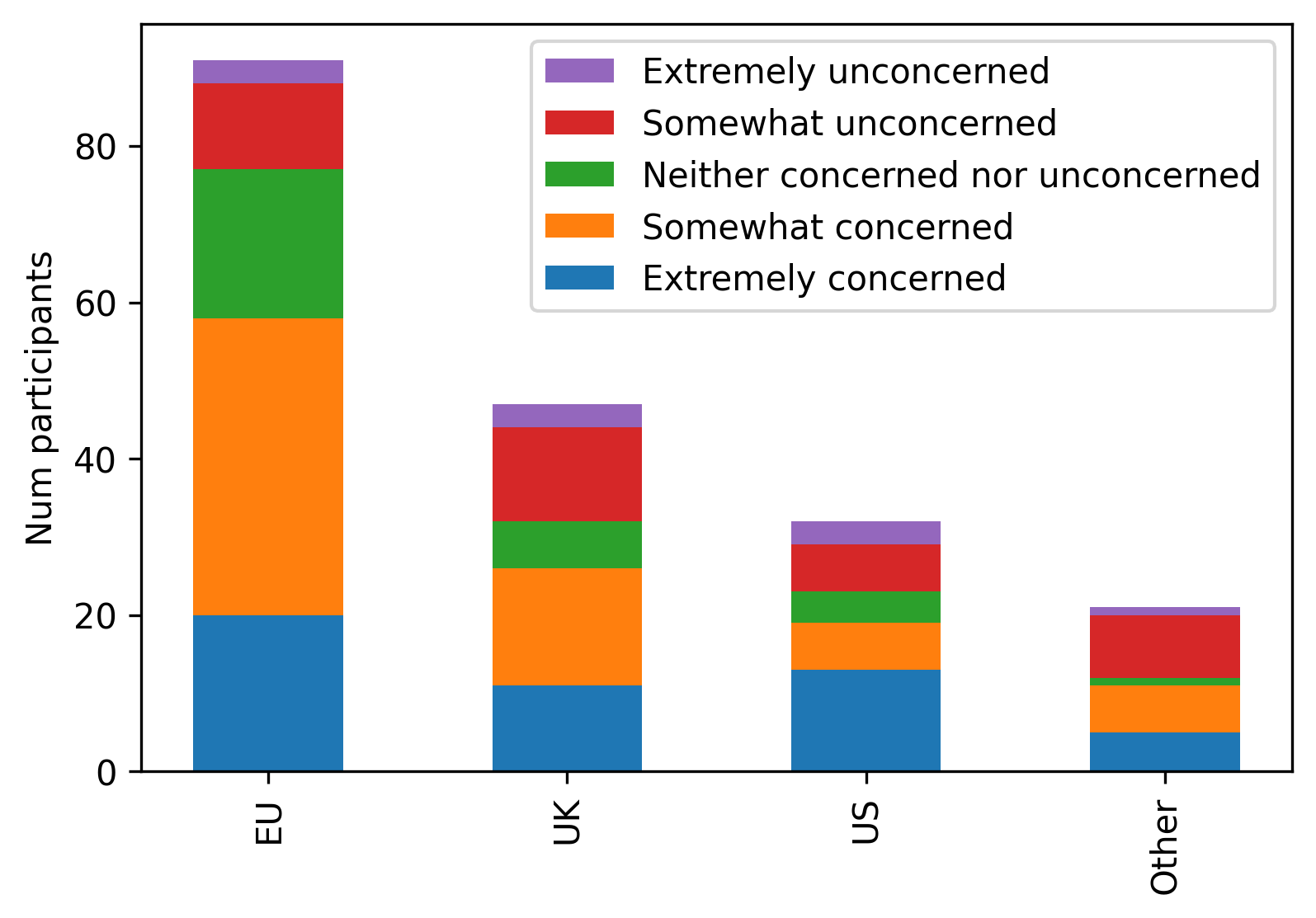}
    \caption{Participants' concern over government data use ``harming [their] personal safety or the safety of those in [their] community''(X68).}
    \label{fig:gov-use-harm}
\end{figure}

Stepping back from asking about a specific data source, we asked participants about their confidence in \textit{general} in their government's handling of any location data marked for COVID-19 tracking. 

Participants largely indicated a lack of trust that their government would use its citizens' location data conservatively. 72\% responded that they thought it was somewhat or extremely unlikely that their government would delete the data (X66; N=190), and 65\% said it was somewhat or extremely unlikely that their government would use it only for the purposes of tracking COVID-19 (X67; N=191). However, 26\% thought it was at least somewhat likely that their government would use the data only for studying and mitigating COVID-19 tracking. 

Participants also indicated concern that such data sharing or collection would be harmful to their safety or the safety of those in their community (X68; N=191) with 26\% saying they were extremely concerned, and 34\% saying they were somewhat concerned. 19\% were somewhat \textit{un}concerned, and only 5\% were extremely unconcerned, as shown in Figure~\ref{fig:gov-use-harm}.

\subsection{Desired app features}

\begin{figure*}
    \includegraphics[scale=1]{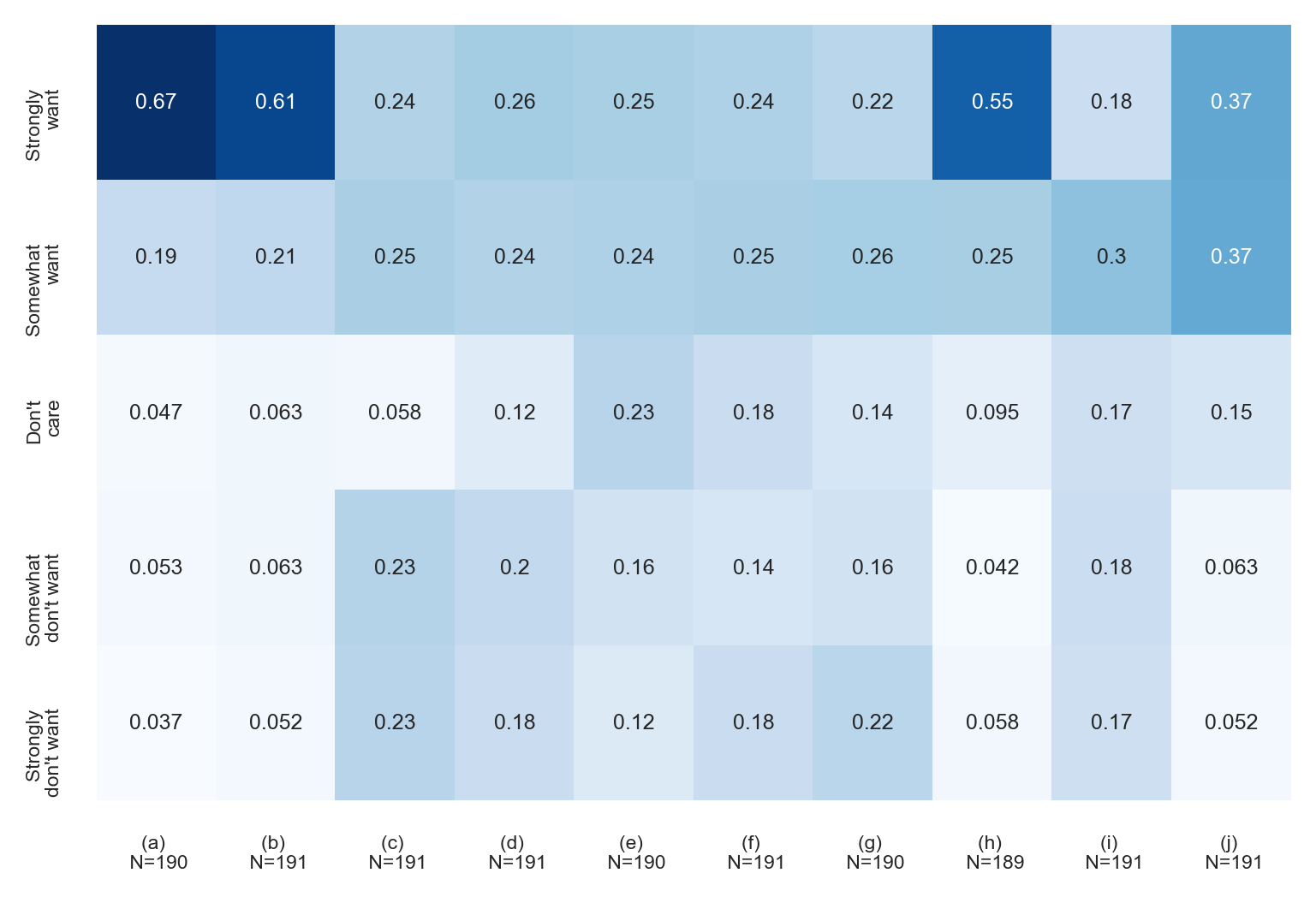}
    \caption{Features desired by participants in a COVID-19 tracking app (X72). Popular features bolded.\newline
    \textit{\textbf{(a) Notify you if you came close to someone who later tested positive for COVID-19\newline
    (b) Notify anyone you came close to in the past two weeks if you tested positive for COVID-19}\newline
    (c) Make your location history for the past two weeks publicly available if you tested positive for COVID-19\newline
    (d) Make public a database of the location histories of anyone who tested positive for COVID-19\newline
    (e) Notify you if your neighbors were not isolating themselves as recommended or mandated\newline
    (f) Let you notify the authorities if you saw people you suspected or knew to be breaking the isolation recommended or mandated\newline
    (g) Automatically notify the authorities if people were not isolating as mandated\newline
    \textbf{(h) Used by scientists to study trends, not individuals\newline}
    (i) Geofence to enforce mandatory or voluntary quarantine\newline
    \textbf{(j) General assessment of social distancing in an area to display areas of high congregation} }}
    \label{fig:section4-features}
\end{figure*}

Participants responded positively to four of the ten potential app features we presented, shown in Figure~\ref{fig:section4-features}: they desire contact tracing notifications (a, b) and general studying of trends (h, j). They are fairly evenly split on other features, even the more privacy-violating features, like (d), in which the app would make public a database of location history for anyone who tested positive for COVID-19, and (g), in which the app would automatically notify the authorities if people were not isolating as mandated.

Interestingly, 49\% of participants were at least somewhat comfortable with feature (c), ``make your location history for the past two weeks publicly available if you tested positive for COVID-19,'' whereas in Section~\ref{section:cell-tower} X15 only 27\% were at least somewhat comfortable with cell tower data being made public if they tested positive. The same difference occurs in the May 1 data. This difference could be due to users having different perceptions of control and autonomy in the two scenarios, with users perceiving (and perhaps having) more control over the behaviors of their apps versus the behaviors of the cellular infrastructure.

\subsection{Cursory analysis of later responses: similar trends; mixed feelings about proximity tracking; negative towards other data sources}

Due to growing conversations about proximity tracking and alternate sources of location data, we added sections to our survey in week 3. Here we present a preliminary analysis of the data from weeks 3, 4, and 5, focusing on the most recent, week 5 (May~1). 

In an initial (non-statistical) comparison with survey data from April 17, 23, and May 1 (weeks 3, 4, and 5 of our survey), many of the same general opinions captured on April 1 and 3 appear. We also present preliminary results about opinions on proximity tracking and the use of surveillance camera footage and credit card data history as location data sources, questions which were also added in week 3.

Geographic, gender, and age distribution for weeks 3, 4, and 5 were similar to week 1: participants were primarily young, from the UK or Europe, and the gender split was fairly even. In week 5 less than 10\% of participants were from the US (lower than other weeks) and participants were 59\% male.

\paragraph{Comparison with April 1 and 3 data.} 

In comparison to Section~\ref{section:likelihood}, on May 1 approximately the same percentage of participants said they would download a new app to study or mitigate COVID-19 ``that protects your data perfectly'' --- 68\% --- but fewer would be likely to download an app that shared their location with their government (24\% somewhat or extremely likely), down from 51\% on April 1 (Figure~\ref{fig:section3_0-how-likely}). April 17 and 23 show similar numbers to May 1, with fewer participants being likely to download a location-tracking app that would share their location with their government. We present these numbers without analysis of statistical significance: this drop could be due to randomness in our samples, or it could indicate less willingness in later dates to download an app if the data would be shared with the government. 

Participants' opinions about cell tower data (Section~\ref{section:cell-tower}), trusted existing apps (Section~\ref{section:existing-app}), and trusted companies and entities (Section~\ref{section:who_develops}) remained similar to those from April 1 and 3, with participants expressing comfort in companies and entities that align with results from April 1 and 3: Universities, Apple, Google, and Microsoft (Microsoft was added as an option after our April 3 survey, and Google and Apple's announcement about creating a contact tracing protocol occurred on April 10~\cite{appleGoogleAnnouncement}).

\paragraph{Proximity tracking.} \label{section:later-results}

\begin{figure}
    \centering
    \includegraphics[scale=1]{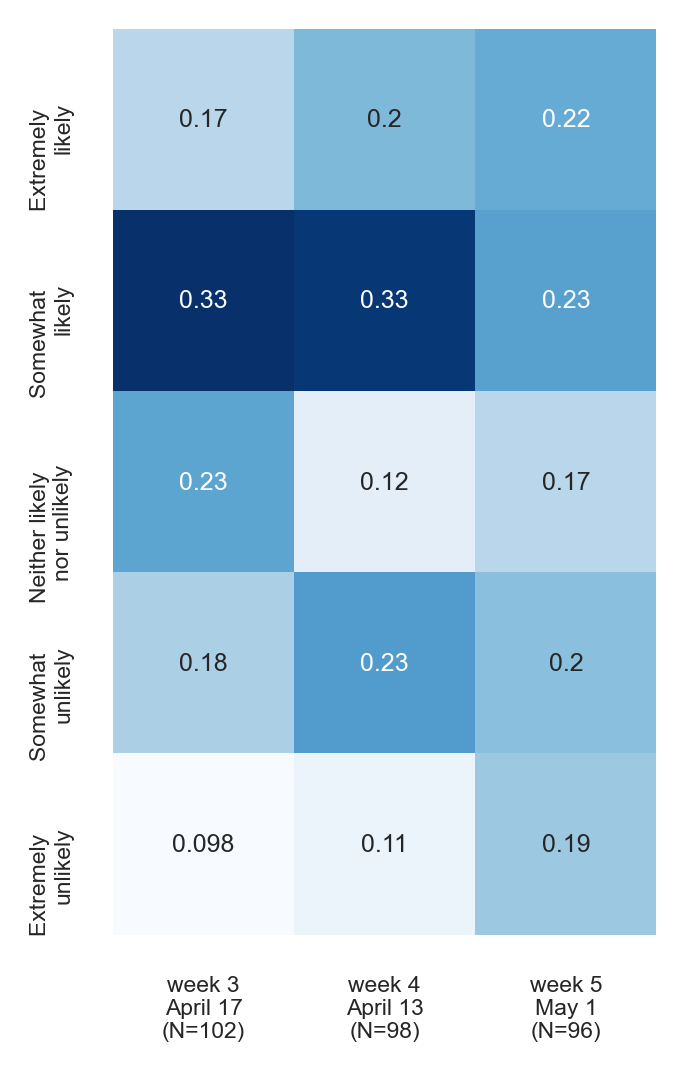}
    \caption{Participant self-reported likelihood of downloading a proximity tracking situation on April 17, 23, and May 1.}
    \label{fig:prox-comparison}
\end{figure}

One new addition after the initial April 1 and 3 survey (in week 3, on April 17) was a section about proximity tracking, as the conversation about proximity tracking has continued to grow amongst the privacy community and the public. 
Because of the significant current public discussions about proximity tracking apps, we focus a sizeable portion of our preliminary analysis on surfacing example qualitative results, which we turn to after our quantitative summaries.

In weeks 3-5, slightly less than half of the participants (46\%-48\%) were somewhat or extremely comfortable with their phone manufacturer or operating system conducting proximity tracking (X121). In weeks 3 and 4, participants were approximately as comfortable with that same data being shared with their government (50\% both weeks); on week 5, 33\% of participants were somewhat or extremely comfortable with proximity data being shared with their government by their phone manufacturer or operating system (X122). This drop may either represent less confidence in government, or it may be due to randomness in our data.

Participants' self-reported likelihood of downloading a proximity tracking app was approximately the same as their likelihood of downloading a location-tracking app in weeks 3 and 4 (X124; figures for download likelihood for all situations and all 5 weeks are available in Appendix~\ref{download-likelihood-figs}). In week 5, proximity tracking received a more mixed response, as shown in Figure~\ref{fig:prox-comparison}. While the mixed response may be either due to sample randomness (more participants from week 5 were from Europe) or declining support for proximity tracking, these results show that, as a whole, the public would not strongly prefer a proximity tracking app or protocol instead of a location tracking app, at least at the time of the survey. 

Initial analysis of qualitative data shows values and concerns in line with those presented in Section~\ref{section:existing-app}: participants reasoned about privacy, security, a ``slippery slope'', and data sharing. Some reasoned that they would prefer proximity tracking, and others, location tracking.

Several participants preferred proximity tracking for security and privacy reasons. W5P42 mentioned adversarial intentions: ``\textit{I think proximity is better than location as it does not track a persons whereabouts, which I think can be manipulated more easily for nefarious things (e.g. stalking). Proximity tracking seems more anonymous and safer in that aspect.}''  W3P4 reasoned that is more private and that additionally it might be more accurate than location tracking: ``\textit{proximity tracking seems better from a privacy perspective. Perhaps even more accurate?}'' For W4P25 and their community, proximity data is less sensitive than location data: ``\textit{It is better. My family and friends are paranoid about the government knowing their whereabouts at all or any given location. This will give more peace to consumers.}''

Some felt that both methods were too privacy invasive. W5P72 wrote ``\textit{They both seem shady to me.}''; W5P12 said ``\textit{Not as intrusive but still terrifying.}'' Others would be begrudgingly willing to install: W3P54 said, ``\textit{Again, I believe I could get used to it if that is what this comes down to.}'' W3P45: ``\textit{It is for a good cause, but the idea of this kind of tracking makes me uncomfortable.}''

Others preferred location tracking. Their reasoning generally fell under the following themes:
\begin{itemize}
    \item Direct and automatic communication between phones is a security risk. W5P73: ``\textit{I’m honestly more comfortable with location tracking than my phone automatically exchanging information with another phone.}'' W5P102: ``\textit{Seems possible for a safety breach between devices sharing data.}'' W3P45: ``\textit{This crosses the line a bit I feel personally, I would want to know that this information is secure and cannot be intercepted.}''
    \item It will be less accurate than location tracking. W5P87: ``\textit{Location tracking would probably be easier to get people on board for as it feels less invasive, however, I think proximity tracking would provide far more information and drastically increase the efficiency of the whole process.}'' 
    \item It is more privacy-invasive than location tracking. W5P88: ``\textit{The government doesn't have to know with who I am.}'' W3P86: ``\textit{It feels more of an invasion of privacy, due to the subtleness of it}.''
    \item Physical security and spoofing. W3P27: ``\textit{It could be used for break-ins (checking if there is someone home)}''. 
\end{itemize}

W5P83 commented the following, showing how the accuracy for either location or proximity tracking apps might be impacted be (perhaps unintentionally) by evasive actions: ``\textit{I believe an app like this is a fallacious system as many people that I know of, at the moment, avoid taking their smartphones outside of the house unless strictly necessary.}''

We are not here to comment on the accuracy of participants' mental models: our results reveal that proximity tracking does not feel more private or secure than location tracking to many potential users. As shown in Section~\ref{section:existing-app}, many participants had similar concerns about location tracking apps. Therefore, developers of both location and proximity tracking apps must consider the various threat and mental models of their potential users, and communicate the threat model to which their protocol or app responses best. W4P41 directly addressed this issue: ``\textit{I am not quite comfortable with this tracking as I don't quite understand the mechanism behind such tracking technology}.'' 

Another overarching theme from Section~\ref{section:existing-app} is that public image matters. Participants reasoned about the app, company, or data source best suited to automated contact tracing on multiple occasions; while the general public may or may not be able to reason accurately about what entities, algorithms, or technical protocols are best suited for contact tracing, it is still driving their decisions.

\paragraph{Other data sources: surveillance camera footage and credit card history.}

\begin{figure}
    \centering
    \includegraphics[scale=1]{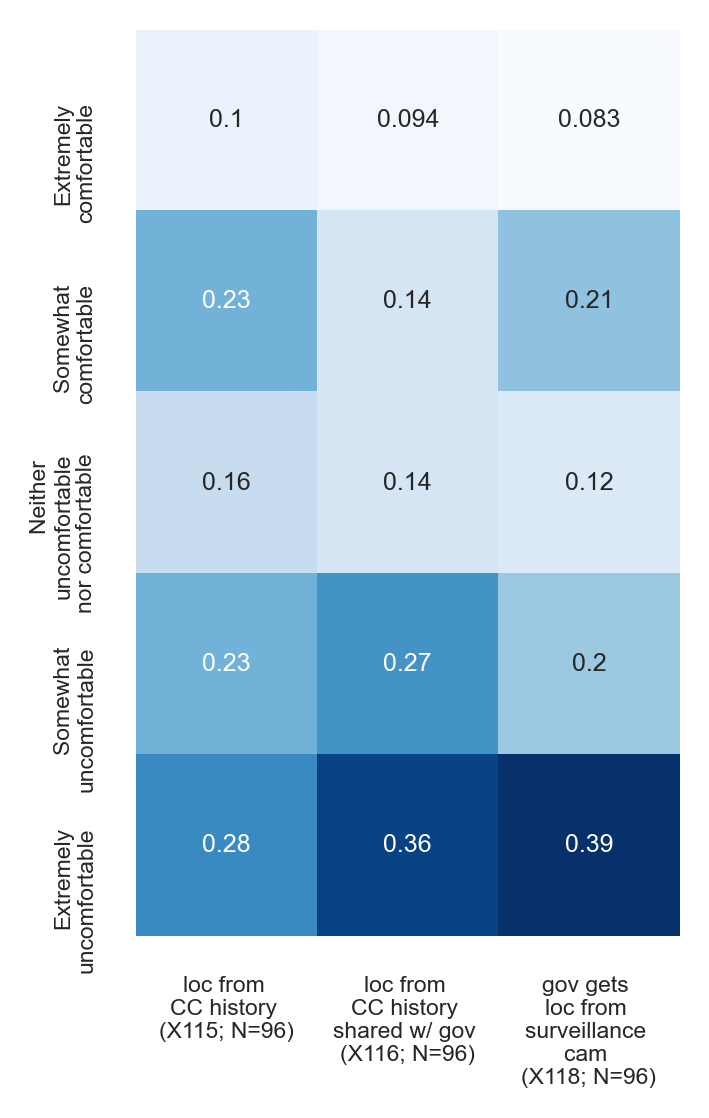}
    \caption{May 1 --- Participant comfort level with their location data being derived from credit card history or surveillance camera footage. }
    \label{fig:other-data-sources-week5}
\end{figure}

Participants were largely uncomfortable with their credit card history being used to derive their location (51\% uncomfortable when not shared with their  government; 63\% uncomfortable when shared with their government), and were less comfortable with the idea of their government deriving their location from surveillance camera footage (69\% uncomfortable), as shown in Figure~\ref{fig:other-data-sources-week5}.

In comparison to cell tower data --- another data source that already exists and does not require a new technology or application --- participants were much less comfortable with credit card history and surveillance camera footage (compare to Figure~\ref{fig:section2_0}; numbers are similar for May 1 participants). W5P14 wrote, ``\textit{It's totalitarian}'', and W5P85 would ``\textit{...find it too similar to a police state situation.}'' However some participants acknowledged that these data sources may already exist: W5P63 wrote that ``\textit{I'm already being watched by surveillance camera footage when I am outside so it would be just utilizing already present stuff.}''

\section{Conclusion}

Here we have presented preliminary results from surveys about public opinion surrounding location privacy during the COVID-19 pandemic. We have been conducting our surveys longitudinally, but report here primarily on our April 1 and 3, 2020 survey results. We have also sampled findings from later runs of the survey. As we have stated throughout the report: these results are preliminary, presented without statistical analysis, and without peer review.

The report adds to other concurrent work about public opinion on potential contact tracing technologies and privacy concerns, and we strongly encourage contact tracing developers, policy makers, and others to consider the user values and concerns presented here, as user cooperation is crucial.

We plan to update this report with more in-depth analysis and longitudinal data over time. Updates will be published at \url{https://seclab.cs.washington.edu/research/covid19/}.

\section{Acknowledgements}

\ifnum\showall=1

We are very thankful to Karl Weintraub for general discussions, sanity, and his Pandas knowledge. We are also thankful to Gennie Gebhart for early discussions about this work. We thank Christine Chen, Ivan Evtimov, Joseph Jaeger, Shrirang Mare, Alison Simko, Robert Simko, and Eric Zeng for their valuable feedback on pilot versions of our survey. Additionally, we are thankful to Gennie Gebhart, Joseph Jaeger, and Elissa Redmiles for their valuable feedback on a draft of this paper. 
We are also very thankful to Prolific for supporting our research by waiving their fees as part of their COVID-19 fee waiver program. 

\fi

This research was supported in part by a University of Washington Population Health Initiative's COVID-19 Rapid Response Grant and by the University of Washington Tech Policy Lab, which receives
support from: the William and Flora Hewlett Foundation, the John D. and Catherine T. MacArthur Foundation, Microsoft, the Pierre and Pamela Omidyar Fund at the Silicon Valley Community Foundation.

{\normalsize \bibliographystyle{acm}
\bibliography{bibliography.bib}}

\ifnum\showall=1
\appendix
\section{Survey Protocol}
\label{survey-protocol}
The survey protocol is below. We give section headings and descriptors for the reader's reference here; participants did not see headers. Unless otherwise specified, all questions were answered on a five-point Likert scale.

The logo of our institution (with the institution name prominent) appear as a header to each survey page. Our lab and department name did not.

On April 10 (week 2), we fixed typos, and added `tax agency' as an option for the government sectors.

On April 17 (week 3) we made the following changes to the survey:
\begin{itemize}
    \item fixed typos;
    \item standardized the use of italics and bold text;
    \item added `Microsoft' as an option in Section~\ref{protocol:new-app};
    \item added a group of questions about other location data sources (Section~\ref{protocol:other-loc});
    \item added a group of questions about proximity tracing (Section~\ref{protocol:prox}). 
\end{itemize}

\subsection{Consent and Screening}

This is a survey about \textbf{location tracking and Coronavirus (COVID-19)} by researchers at the University of Washington, in Seattle, Washington, USA. The University of Washington’s Human Subjects Division reviewed our study, and determined that it was exempt from federal human subjects regulation. We do not expect that this survey will put you at any risk for harm, and you don’t have to answer any question that makes you uncomfortable. In order to participate, you must be at least 18 years old, regularly use a smartphone, and able to complete the survey in English. We expect this survey will take about 15-20 minutes to complete. 

If you have any questions about this survey, you may email us at [study-specific-email].

Thanks for taking our survey! To start, please answer the two questions below...

Are you at least 18 years old? [yes, no]

Do you use a smartphone regularly? [yes, no]

\subsection{Demographics I}

This survey involves questions about \textbf{COVID-19}, the disease caused by SARS-CoV-2 (commonly known as \textbf{coronavirus}).

X6: How concerned are you about COVID-19? 

X7: Do you believe that social distancing is an important tool for slowing the spread of COVID-19? [yes, no]

X8: Averaged over the past week, approximately how many hours much time per day did you spend out of your home, within 6 feet (2 meters) of other people? (e.g., getting groceries, working at an essential job like in a hospital, in a grocery store, etc). 
[`I did not leave my home', 0-1 hours per day, 2-3 hours per day, ... , 7-8 hours per day, 8+ hours per day] 

X9: In which country do you currently reside? [drop-down country list]

X10: For respondents in the USA: in which state do you currently reside? [drop-down US state list]

\subsection{Cell phone manufacturer and provider location data}

\textit{Cell phone manufacturers and cellular providers have access to your physical-world location.}

X12: How comfortable are you with your cell phone manufacturer or your cellular carrier using your location data for the purposes of studying or mitigating the spread of COVID-19? 

X13: How comfortable are you with your cell phone manufacturer or your cellular carrier sharing your location data for the past two weeks \textbf{with your government} for the purposes of studying or mitigating the spread of COVID-19? (\textbf{Regardless of whether you test positive for COVID-19.})

X14: \textbf{If you tested positive for COVID-19}, how comfortable would you be with your cell phone manufacturer or your cellular carrier sharing your location data for the past two weeks \textbf{with your government} for the purposes of studying or mitigating the spread of COVID-19?

X15: \textbf{If you tested positive for COVID-19}, how comfortable would you be with your cell phone manufacturer or your cellular carrier sharing your location data for the past two weeks \textbf{publicly}?

X16: Optionally, do you have any other thoughts about your cell phone manufacturer or your cellular carrier sharing your location data for the purposes of studying or mitigating the spread of COVID-19? [free response]

\subsection{Existing app location data}
\textit{Some phone applications have access to your physical-world location, either when the application is in use or all the time. \textbf{Suppose the makers of an existing app on your phone started using your GPS location data to study or mitigate the spread of COVID-19.} For example, this could include disclosing past locations of known positive COVID-19 cases to the public or to the government, or alerting people who have crossed paths with the positive case.}

X18- Below we've listed 15 commonly-used apps. For the apps that you use regularly: how comfortable are you with the following apps using your location data for the purposes of studying or mitigating the spread of COVID-19? [``I don’t use this app'' + 5-point Likert scale for each of the following apps]

\begin{itemize}
    \item Google Maps
    \item Apple Maps
    \item Waze
    \item Facebook
    \item Instagram
    \item TikTok
    \item WhatsApp
    \item Facebook Messenger
    \item Zoom
    \item Uber
    \item Lyft
    \item Airbnb
    \item Calorie Counter (MyFitnessPal)
    \item FitBit
    \item AllTrails
\end{itemize}

\textit{Suppose that one of the apps that you regularly use  -- not necessarily one of the ones above -- started using your location data to study or mitigate the spread of COVID-19. }

X20: How comfortable are you with this app using your location data for the purposes of studying or mitigating the spread of COVID-19? 

X22: \textbf{If you tested positive for COVID-19}, how comfortable would you be with this app sharing your location data for the past two weeks \textbf{publicly}?

X23: Consider all the apps you regularly use on your phone (not just the apps listed earlier). Which app would you \textbf{most} trust to use your location data for the purposes of studying and mitigating COVID-19? Why? [free response]

X24: Which app that you currently use would you \textbf{least} trust to use your location data for the purposes of studying and mitigating COVID-19? Why? [free response]

\subsection{Current use of COVID-19 app}

X25: Have you used any apps that help track the spread of COVID-19? (i.e. Singapore's ``TraceTogether'') [yes, no]

\underline{If yes, participants branch to `already have app.'}

\underline{If no, participants continue.}

\subsection{New app, perfect privacy}

\textit{Imagine there is a \textbf{new} app that would track your location at all times for the purposes of mitigating the spread of COVID-19.}
 
\textit{Suppose that this app protects your data perfectly.}
 
X50: How likely would you be to install and use this app?

X51: Would this app change your current behavior?

X52: Optionally, please use this space tell us any initial thoughts you have about such an app. [free response]

\subsection{New app, app makers know location}
\textit{Imagine there is a new app that would track your location at all times for the purposes of studying or mitigating the spread of COVID-19.} \label{protocol:new-app}

\textit{Suppose now that the makers of the application would know your location at all times, but would not share your location with any other entity. }
 
X55: How likely would you be to install and use this app?

X56: Now, suppose that the app is made by one of the following companies, all of which already have created popular apps. Please rate how comfortable you would be if each company were responsible for this new app. [``I don't know enough about this company to make a  decision'' + 5-pt Likert scale for each of the following]
\begin{itemize}
    \item Google (Google Maps, Waze, etc)
    \item Apple (Apple Maps)
    \item Facebook (Facebook, Facebook Messenger, Instagram, WhatsApp)
    \item Microsoft (Skype, OneDrive, etc)\footnote{Added April 17 (week 3).}
    \item ByteDance (TikTok)
    \item Zoom Video Communication
    \item Uber
    \item Lyft
    \item AirBnb
    \item MyFitnessPal
    \item AllTrails
    \item FitBit
\end{itemize}

X57: Suppose that the app is made by one of the following general entities. Please rate how comfortable you would be if one of the following were responsible for this new app, which would use the location data they collect from your smartphone to track the spread of COVID-19. [5-pt Likert scale for each of the following]
\begin{itemize}
    \item A university research group
    \item An activist group 
    \item An industry startup
    \item Your government
    \item The United nations
\end{itemize}

X58: Optionally, please use the space below to elaborate on your thoughts about one or more companies using your location data for the purposes of tracking COVID-19. [free response]

\subsection{New app, app makers share data with government}

\textit{Again, imagine there is a new app that would track your location at all times for the purposes of studying or mitigating the spread of COVID-19.}
 
\textit{Suppose now that the makers of the application would know your location at all times, and would also share that data with your government if you were diagnosed with COVID-19.}

X61: How likely would you be to install and use this app?

X62: If the government’s use of the data were \textbf{supervised by a judge}, how likely would you be to install and use this app?\footnote{This question, and the rest of this section, was added on April 17 (week 3) as a previous version was ambiguous.}

\textit{Now suppose that the makers of the application would share your location data with your government \textbf{only if you tested positive for COVID-19.}}

X112: How likely would you be to install and use this app?

X64: Optionally, do you have any other thoughts about a company that is doing COVID-19 tracking sharing your location with your government? [free response]

X115:  Optionally, do you have any other thoughts about judicial oversight of the government’s usage of location data? [free response]

\subsection{Other location data sources: Credit card history and surveillance camera footage}\label{protocol:other-loc}

\textit{There are other ways to track someone's location. One is the use of video cameras in public places. Another is the use of credit card purchasing histories.}\footnote{This section added April 17 (week 3)} 

X115: How comfortable would you be with your \textbf{credit card company} deriving your location history for the past two weeks for the purposes of studying and mitigating the spread of COVID-19?

X116: How comfortable would you be with your \textbf{credit card company} deriving your location history for the past two weeks \textbf{and sharing it with your government} for the purposes of studying and mitigating the spread of COVID-19?

X117: Optionally, do you have any other thoughts about your location history being derived from your \textbf{credit card} purchase history?

\subsection{Proximity tracing}
\label{protocol:prox}
\textit{One alternative to location tracking for the purposes of studying or mitigating COVID-19 is \textbf{proximity tracing}, in which your phone would automatically exchange information with every phone within 6 feet (2 meters) of your phone, \textbf{keeping track of your close physical encounters, but not tracking your actual location}. This data could then be used to reconstruct your close encounters if you contracted COVID-19, or could alert you if someone you had been in close physical proximity to tested positive for COVID-19.}\footnote{This section added April 17 (week 3)} 

X121: Imagine that your \textbf{cell phone manufacturer or phone operating system} would conduct proximity tracing for the purposes of studying or mitigating COVID-19 (and, vice versa, other phones will record that they have been in the proximity of your phone). How comfortable would you be with this?

X122: Suppose that your \textbf{cell phone manufacturer or phone operating system} would share this proximity data \textbf{with your government} if you tested positive for COVID-19. How comfortable would you be with this?

X123: Optionally, do you have any other thoughts about your cell phone manufacturer or phone operating system tracking other phones nearby? [free response]

\textit{Imagine instead there is a new \textbf{app} that would conduct proximity tracing for the purposes of studying or mitigating COVID-19: that is, it would not track your location, but would instead keep track of other phones that you are nearby (and, vice versa, other phones with this app will record that they have been in the proximity of your phone).}

X124: How likely would you be to download this app?

X125: Now, suppose that the proximity tracing app is made by one of the following companies. Please rate how comfortable you would be if each company were responsible for this new app.
[``I don't know enough about this company to make a  decision'' + 5-pt Likert scale for each of the following]
\begin{itemize}
    \item Google (Google Maps, Waze, etc)
    \item Apple (Apple Maps)
    \item Facebook (Facebook, Facebook Messenger, Instagram, WhatsApp)
    \item Microsoft (Skype, etc)
    \item ByteDance (TikTok)
    \item Zoom Video Communication
    \item Uber
    \item Lyft
    \item AirBnb
    \item MyFitnessPal
    \item AllTrails
    \item FitBit
\end{itemize}

X126: Now, suppose that the proximity tracing app is made by one of the following general entities. Please rate how comfortable you would be if each entity were responsible for this new app.
\begin{itemize}
    \item A university research group
    \item An activist group 
    \item An industry startup
    \item Your government
    \item The United nations
\end{itemize}

X127: Optionally, do you have any other thoughts about an app that tracks other phones nearby?

X128: Optionally, do you have any other thoughts about proximity tracking versus location tracking for the purposes of studying or mitigating COVID-19?

\subsection{Government use of data}
\textit{\textbf{If the government acquired your location data or proximity data\footnote{'or proximity data' added April 17}}(i.e. from an app on your phone, from your cell phone carrier, etc) \textbf{for the purposes for studying and mitigating COVID-19}....}
 
X66: How likely do you think it is that your government would \textbf{delete} the data after the pandemic ends?

X67: How likely do you think it is that your government would \textbf{only} use the data for the purposes of tracking COVID-19?

X68: How concerned would you be about your government’s use of your location data harming \textbf{your personal safety} or the safety of those in your community? 

X69: Suppose your location was shared with only a specific sector of the government. For each of the following sectors of government, please rate how comfortable you would be with them having access to your location data. 
\begin{itemize}
    \item Federal Disease Tracking Agency (US: CDC)
    \item Your state or City Health Department
    \item Tax Agency (US: IRS)
    \item Local law enforcement (state, country, city, etc)
    \item Immigration authorities (US: CBP or ICE)
\end{itemize}

X70: Optionally, please use the space below to elaborate on your thoughts about the government having access to your location data for the purposes of COVID-19 tracking.  [free response]

\subsection{App features}
\textit{In some countries, such as South Korea, China, and Singapore, there do exist apps to monitor the spread of COVID-19 through location tracking. These apps can have multiple purposes, including:}

- \textit{Alerting the user if they have come into contact with someone who later tests positive with COVID-19;}

- \textit{Helping the community or law enforcement enforce isolation and quarantine edicts;}

- \textit{Tracing viral strains through the community.}
 
X72: If a new app were deployed in your country to mitigate the spread of COVID-19, which of the following features would you want it to have? (5-point Likert-scale for each of the following:)
\begin{itemize}
    \item Notify you if you came close to someone who later tested positive for COVID-19
    \item Notify anyone you came close to in the past two weeks if you tested positive for COVID-19
    \item Make your location history for the past two weeks publicly available if you tested positive for COVID-19
    \item Make public a database of the location histories of anyone who tested positive for COVID-19
    \item Notify you if your neighbors were not isolating themselves as recommended or mandated
    \item Let you notify the authorities if you saw people you suspected or knew to be breaking the isolation recommended or mandated
    \item Automatically notify the authorities if people were not isolating as mandated
    \item Used by scientists to study tends, not individuals
    \item Geofence to enforce mandatory or voluntary quarantine
    \item General assessment of social distancing in an area to display areas of high congregation
\end{itemize}

X73: Optionally, do you have any other thoughts about what you would want such an app to do? [free response]

X74: Optionally, do you have any other thoughts about what you would want such an app to NOT do? [free response]
 
X75: Is such an app available in your country? [Yes / No / I’m not sure]
\underline{If yes, participants branch to ‘App available’}

\subsection{Section 5.0: Prior privacy preferences}

\textit{We’re now going to ask you about your thoughts about location sharing with your government BEFORE COVID-19.}
 
80 - In Oct 2019 (before the first known cases of COVID-19), how comfortable would you have been with your location data being shared with the government in general?

81 - In Oct 2019 (before the first known cases of COVID-19), how comfortable were you with your location data being shared with the following sectors of government? [5-Point Likert scale for each of the following:]
\begin{itemize}
    \item Federal Disease Tracking Agency (US: CDC)
    \item Your state or City Health Department
    \item Tax Agency (US: IRS)
    \item Local law enforcement (state, country, city, etc)
    \item Immigration authorities (US: CBP or ICE)
\end{itemize}

82 - Optionally, please use the space below to elaborate on your thoughts about one or more companies sharing your location data with some part of the government (before COVID-19). [free response]

\subsection{Demographics II}

Almost done!

X39: What is your age? (you may answer approximately if you do not know, or wish not to say exactly) [free response]

X40: What is your gender identity? [free response]

X41: What political party do your views typically align with? [free response]

X42: What are your top three news sources? (i.e. Twitter, Facebook, Fox News, CNN, NPR, New York Times, etc) [free response]

X43: Regarding COVID-19, are you in high risk group or live with someone with high risk? [yes / no]

X44: Are you generally interested in or concerned about privacy and technology? [yes / no]

X45: Do you know how to change location permissions for apps on your phone? [yes / no]

\subsection{Branch: app is available}

X76: Have you downloaded the app in your country to mitigate or study the spread of COVID-19? [yes, no]

X77: If you have not downloaded the app: why not? what changes, or assurances by the manufacturer or government (if any) would you want to see to the app before downloading? [free response]

X78: What are your thoughts about the privacy properties of this app? [free response]

\subsection{Branch: Already have app}

You indicated that there is an app (or apps) available in your country to track, study, or mitigate COVID-19. This section will ask about that app.

X27: What is the name of the app, or apps?

X28: Why did you install and use it?

X29: Do you know anyone who did not download the app? [yes, no]

X30: If so, why did they not install it?

X31: What concerns, if any, do you have about the app?

X32: If you had or have concerns, what outweighed the concerns and lead you to the decision to download the app?

X33: What concerns, if any, do you have about your government having access to your location data?

X34: What concerns, if any, do you have about the app makers having access to your location data?

X35: Do you expect the app makers to stop storing your location data after the pandemic is over?

X36: Do you plan to delete the app after the pandemic?

X37: Anything else you’d like to say about the app and/or your concerns?

\section{Participants' Download Likelihood of Contact Tracing Apps: Weeks 1-5}
\label{download-likelihood-figs}

This appendix shows the download likelihood data in each situation for Weeks 1-5.

Week 1:

    \includegraphics[scale=1]{figures/initial_heatmaps/prolific_week1/section3_how_likely_paper.png}

\newpage

Week 2: 

    \includegraphics[scale=1]{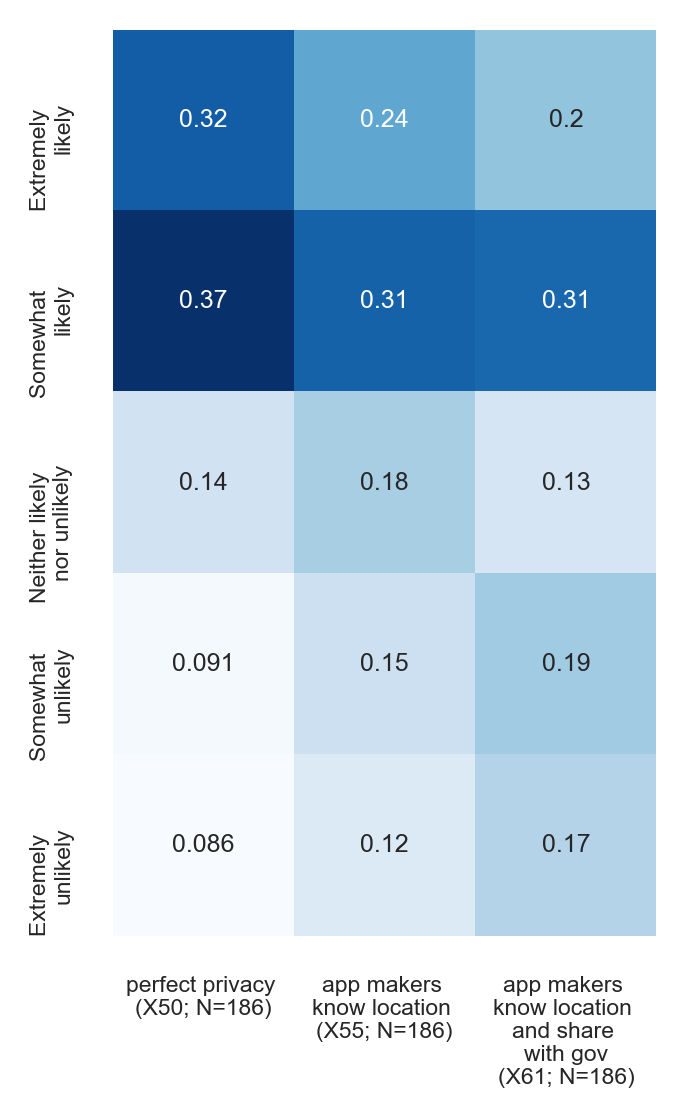}

Week 3:    

    \includegraphics[scale=.8]{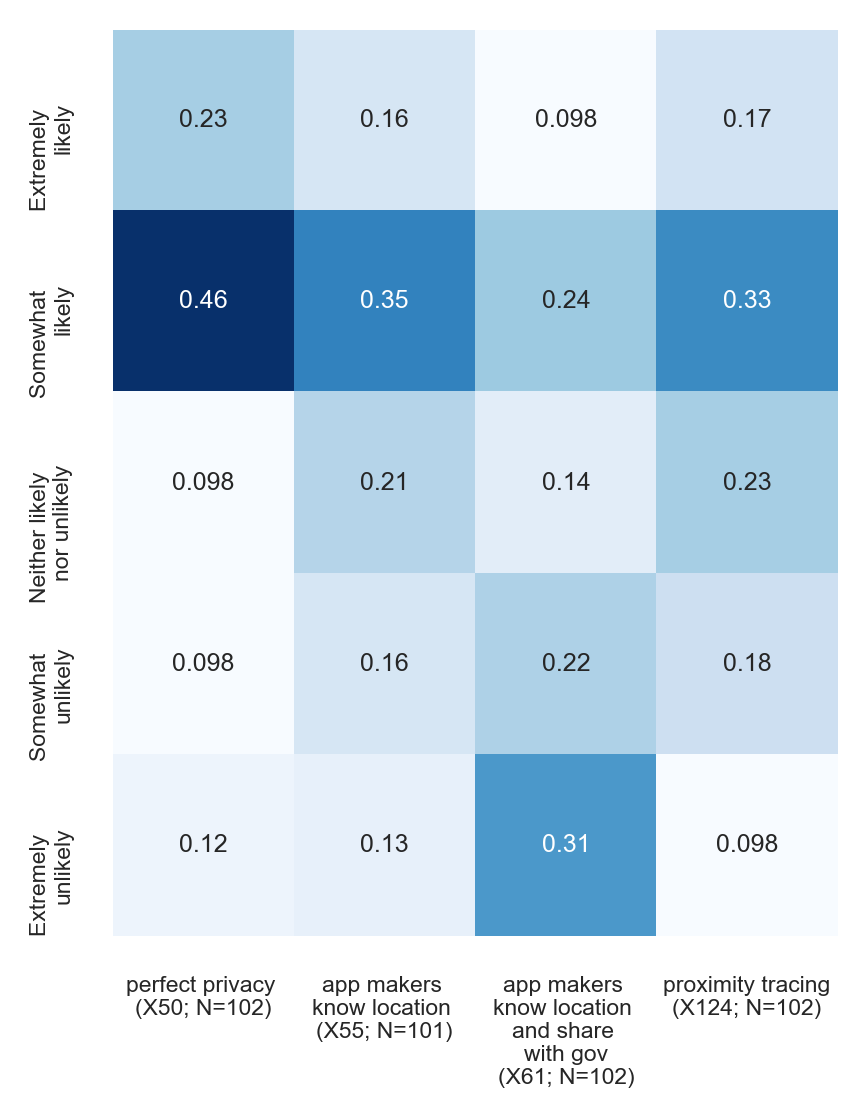}
    
\newpage

Week 4:

    \includegraphics[scale=.8]{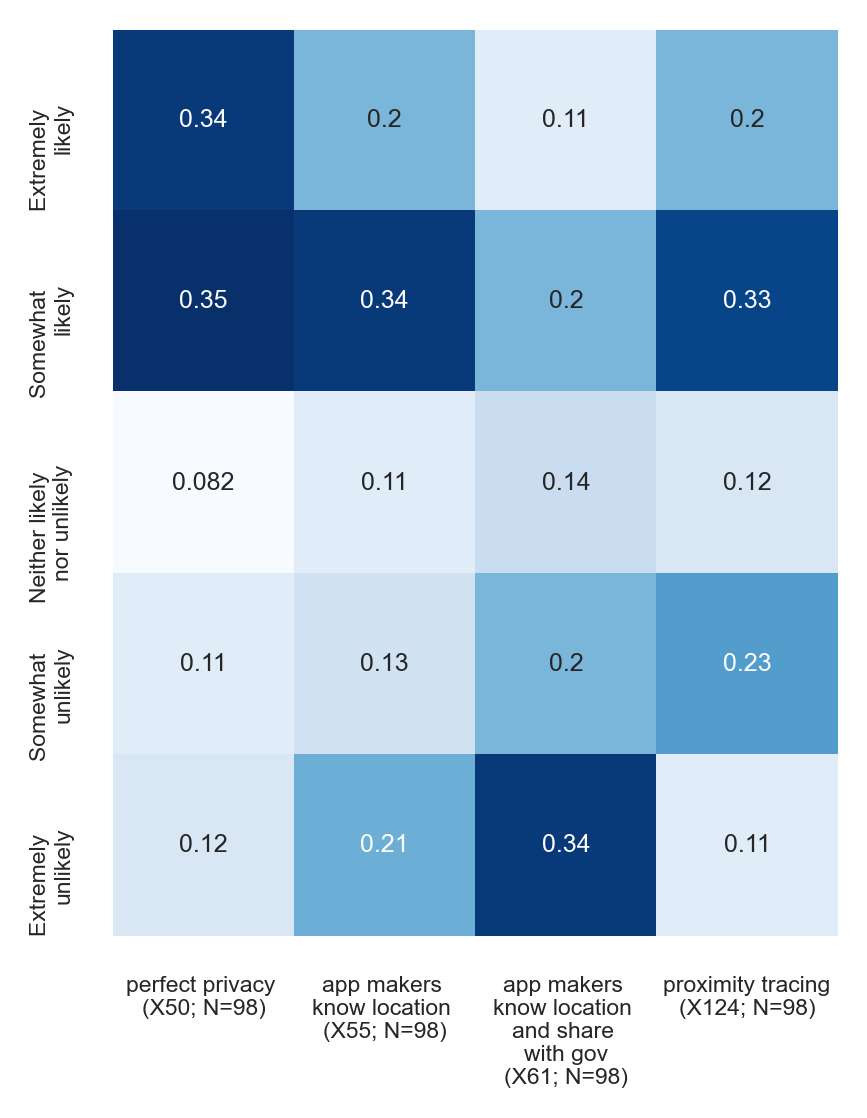}

Week 5:

    \includegraphics[scale=.8]{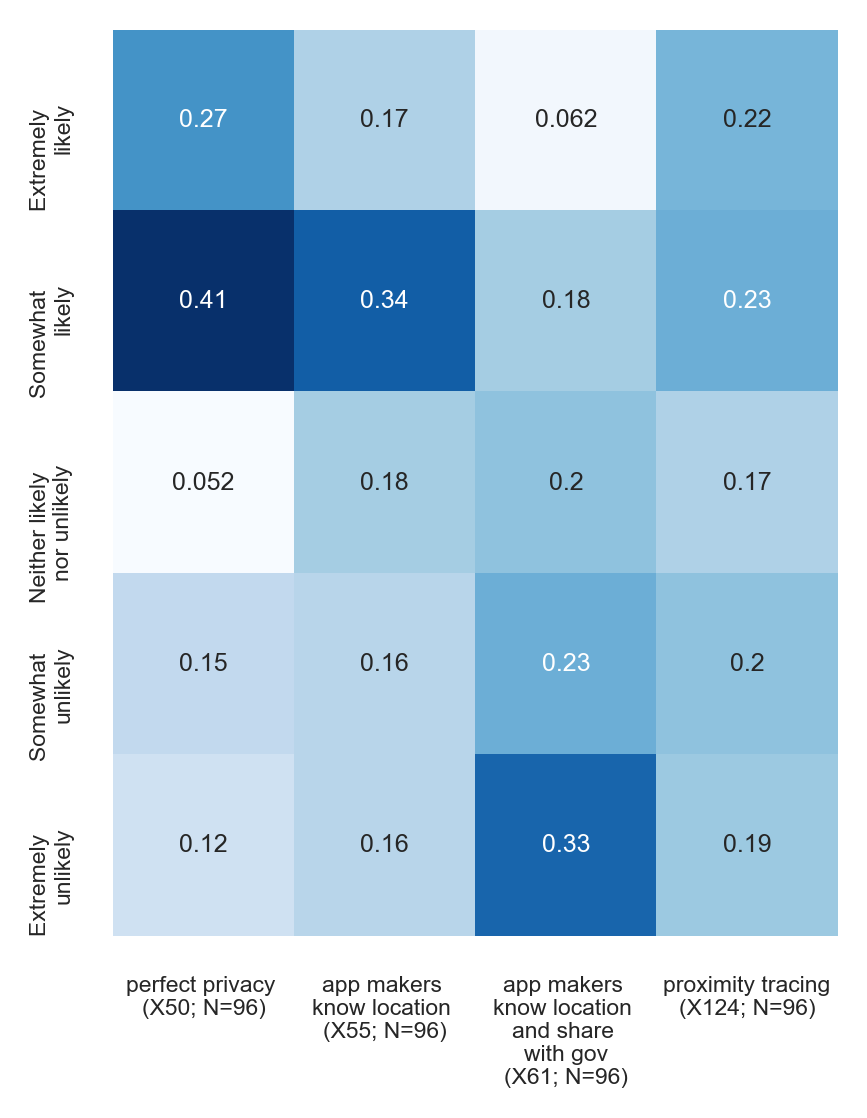}

\fi 

\end{document}